\pdfoutput=1

\documentclass[pra,10pt,twocolumn,superscriptaddress,footnoteinbib]{revtex4}
\usepackage{amsmath}
\usepackage{latexsym}
\usepackage{amssymb}
\usepackage{graphics,epstopdf}
\usepackage[colorlinks=true, citecolor=blue, urlcolor=blue ]{hyperref}
\usepackage{epsf,graphics,graphicx}
\usepackage{color}

\newcommand{\ed}{\end{document}}
\newcommand{\beq}{\begin{equation}}
\newcommand{\eeq}{\end{equation}}

\date{\today}
\begin{document}

\title{Particle current rectification in a quasi-periodic 
double-stranded ladder}

\author{Madhumita Saha}

\affiliation{Physics and Applied Mathematics Unit, Indian Statistical
Institute, 203 Barrackpore Trunk Road, Kolkata-700 108, India}

\author{Santanu K. Maiti}

\email{santanu.maiti@isical.ac.in}

\affiliation{Physics and Applied Mathematics Unit, Indian Statistical
Institute, 203 Barrackpore Trunk Road, Kolkata-700 108, India}

\begin{abstract}

We report transport properties and particle current rectification operation 
in a double-stranded tight-binding ladder network within non-equilibrium 
Green's function (NEGF) formalism that can easily be generalized in 
multi-stranded 
systems. First, we explore the rectification operation considering the model 
of an artificial double-stranded DNA (dsDNA) molecular system 
with Fibonacci type substitutional sequence. Substitutional sequences form 
quasi-periodic potentials. This analysis may shed new light in designing 
efficient nanoscale rectifier. It can also be directly implemented to 
different biomolecular systems like nucleic acids and most proteins as 
they follow quasi-periodic orders. Motivated with this fact, we consider 
different configurations depending on the choices of on-site energies 
and/or inter or intra strand nearest-neighbor hopping integrals in the 
form of Aubry-Andr\'{e}-Harper (AAH) model (that also obeys quasi-periodic 
order), and in all the cases we find almost $100\%$ current rectification even 
at very low bias region. Along with this, we observe that the phase (positive 
or negative) of rectification can suitably be engineered by tuning the Fermi 
energy and AAH phase. The effects of electron-electron (e-e) interaction and 
temperature are also studied, which show that a reasonably large rectification 
can be observed even for moderate temperature range and e-e interaction 
strength.

\end{abstract}

\maketitle

\section{Introduction}

Molecular systems will be the ultimate functional elements in fabricating
electronic circuits at nanoscale level~\cite{pr1,pr2,pr3}, suppressing 
the use of conventional conducting and semiconducting materials. The 
proposition of designing molecular diode or rectifier given by Ratner and
co-workers essentially triggered the use of single molecules as active 
functional elements~\cite{pr4}. Later, some other theoretical proposals 
have been put forward describing the rectification action using simple
and complex molecular systems~\cite{pr5,pr6,pr7,pr8}. However, a single 
molecular rectifier was first experimentally realized~\cite{pr9} in $2005$ 
with modest rectification ratio (RR) $\sim 10$. Since then the interest 
in this subject is rapidly growing up with considerable experimental and 
theoretical works~\cite{pr10,pr11,pr12,pr13,pr14,pr15,pr16,pr17,pr18,pr19,
pr20,pr21,pr22,pr23,pr24,pr25,pr26,pr27,pr28}. Some of them are 
based on intrinsic properties~\cite{pr9,pr10,pr11,pr12} of the molecules, 
while the others have considered the effects of molecular coupling~\cite{pr13,
pr14,pr15,pr16}. Instead of these significant advances, the present molecular 
diodes have had very limited applications especially due to their low 
conductances, low rectification ratio, high degree of sensitivity to junction 
configuration and much higher operating voltage~\cite{pr9,pr10,pr12,pr13}. 
To circumvent these factors, few other works have been carried 
out~\cite{pr25,pr26,pr27,pr28} in order to get tunable rectification via 
gate voltage, environmental control, etc., though a successful conclusion 
is still lacking and thus further probing is undoubtedly required.

In the last few decades DNA molecules have attracted significant attentions 
in the fields of molecular physics and molecular spintronics due to their
unique and diverse characteristic features~\cite{dna1,dna2,dna3,dna4,dna5,dna6}. 
Using DNA-coralyne complex molecule as a functional element 
Guo {\em et al.}~\cite{dna6} have shown that efficient rectification can be 
performed, and the rectification ratio can be reached up to $\sim 15$ even at 
relatively small bias voltage ($1.1\,$V). Artificial DNA sequences, those are 
usually constructed in a
quasi-periodic manner, even also play appealing roles on transport phenomena,
and several important aspects have already been revealed along this line 
considering different aperiodic lattices like Copper-mean, Nickel-mean, 
Fibonacci~\cite{quasiall}, Thou-Morse~\cite{tm}, and to name a few~\cite{dna5}. 
Appreciating the enormous possibilities of DNA molecule to become a role model 
in future nanoelectronic devices, can we now think about a DNA device where 
large asymmetry in current-voltage curves can be obtained and tuned externally? 
It has been established by us that a $1D$ chain 
with quasi-periodic potential can show strong rectification~\cite{quasi-recti}. 
Naturally, it raises an obvious question that whether an artificial dsDNA which 
follows discrete quasi-periodic sequence can exhibit strong rectification. 
Fermi energy $E_F$ is the only externally tunable parameter for this system 
which can be controlled by gate voltage. Now, in the search of other possible
tuning parameters in an artificial DNA device, we consider a double-strand 
ladder network with quasi-periodic modulations~\cite{macia06,rudo07,ladder1,
ladder2,moura,sousa,ladder3} in the form of well-known Aubry-Andr{\'e}-Harper 
(AAH) model~\cite{aubry1}. Recently AAH model has got significant attention in 
both theoretical and experimental research~\cite{aubry2,aubry3,expt0,expt1,
expt2,expt3,aubry4,aubry5,aubry6,aubry7,aubry2d,aubry2d2,aubry2d3}. 
The quasi-periodicity can be tuned
with the help of AAH phase(s)~\cite{aubry5} and the modulations can be 
incorporated in the diagonal and/or off-diagonal parts.
Thus, in their unique characteristics, AAH models make them suitable 
candidates for designing efficient nanoscale devices, and in the present 
work, we explore how rectification operation can be performed considering 
a double-strand AAH ladder (Fig.~\ref{mod}) with $E_F$ and the AAH phase. 

The primary motivations of this work are as follows. First, we start with an 
artificial dsDNA molecular system with Fibonacci sequence to explore particle current 
rectification operation. If the rectification performance is good enough then 
we can argue that a double-stranded ladder with quasi-periodic sequence will 
be a functional element for designing nanoscale rectifiers since DNA molecule 
can be modeled nicely with quasi-periodic sequences, as already reported in many 
works~\cite{dna5,macia06}. Following the DNA results, we attempt to propose 
a suitable model where much higher rectification can be obtained, and at the 
same time, the rectification ratio can be tuned externally by more parameters 
other than the gate voltage. In our model, there are two phase factors associated 
with the cosine modulations in diagonal as well as off-diagonal parts of the AAH 
Hamiltonian, which will play the key roles for effective controlling. We believe 
that rectification operation in quasi-periodic ladder network is quite interesting, 
and have not been discussed so far, to the best of our concern.

The essential mechanism of rectification relies on the fact that in presence 
of finite bias, a voltage drop takes place along the ladder. Therefore,
average density of states (ADOS) and transmission spectrum become voltage 
dependent. The quasi-periodic potential leads to a gapped and fragmented 
spectrum~\cite{fragment1,fragment2} and breaks the spatial reflection symmetry. 
It makes quasi-periodicity a key ingredient for occurrence of large particle 
current rectification~\cite{quasi-recti}. In a recent 
work~\cite{quasirec1}, strong energy current rectification has been shown 
in a quadratic bosonic system, subjected to quasi-periodic potential, 
which is connected to spin baths having differing temperatures. Within a 
tight-binding (TB) framework we investigate transport properties and particle 
current rectification operation using
the Green's function formalism~\cite{green1,green2,green3,green4,green5,
green6}. First, we investigate strong rectification in an 
artificial dsDNA molecule with Fibonacci sequence which forms quasi-periodic 
potential. The mechanism behind the rectification is solely due to the 
quasi-periodicity and it is completely different from the previous work on 
DNA molecule done by Guo and co-workers~\cite{dna6}. Then, different cases 
are taken into account depending on the specific choices of site energies 
and/or nearest-neighbor hopping (NNH) integrals in the form of AAH modulations. 
In all the cases high degree of rectification is obtained, which on one hand 
appears at sufficiently low bias, and, on the other hand, sustains for a 
wide range of parameter values. We find that the RR can be tuned significantly 
by modulating the AAH phase, keeping the $E_F$ constant. This is one of the key 
advantages of AAH model over other discrete quasi-periodic systems. We also discuss the effects 
of electron-electron (e-e) interaction, treated at the Hartree-Fock (HF) 
mean-field (MF) level~\cite{inter1,inter2,inter3,inter4}, and 
temperature. We find that the asymmetry in current-voltage curves persists over 
a moderate temperature range and for different values of interaction strength. 
We hope that the present analysis can be verified in future with suitable 
experimental setup.

The rest of the work is organized as follows. In Sec. II we present the 
model, TB Hamiltonian and theoretical framework for the calculations. 
The results are thoroughly analyzed in Sec. III, and, few contemporary parts are
discussed in Appendix~\ref{aa} and Appendix~\ref{bb}. Finally, we conclude our 
findings in Sec. IV.

\section{Model, TB Hamiltonian and the Theoretical framework}

\subsection{Model and the Hamiltonian}

\begin{figure}[ht]
\centering
\includegraphics[height=2.75cm, width=8cm]{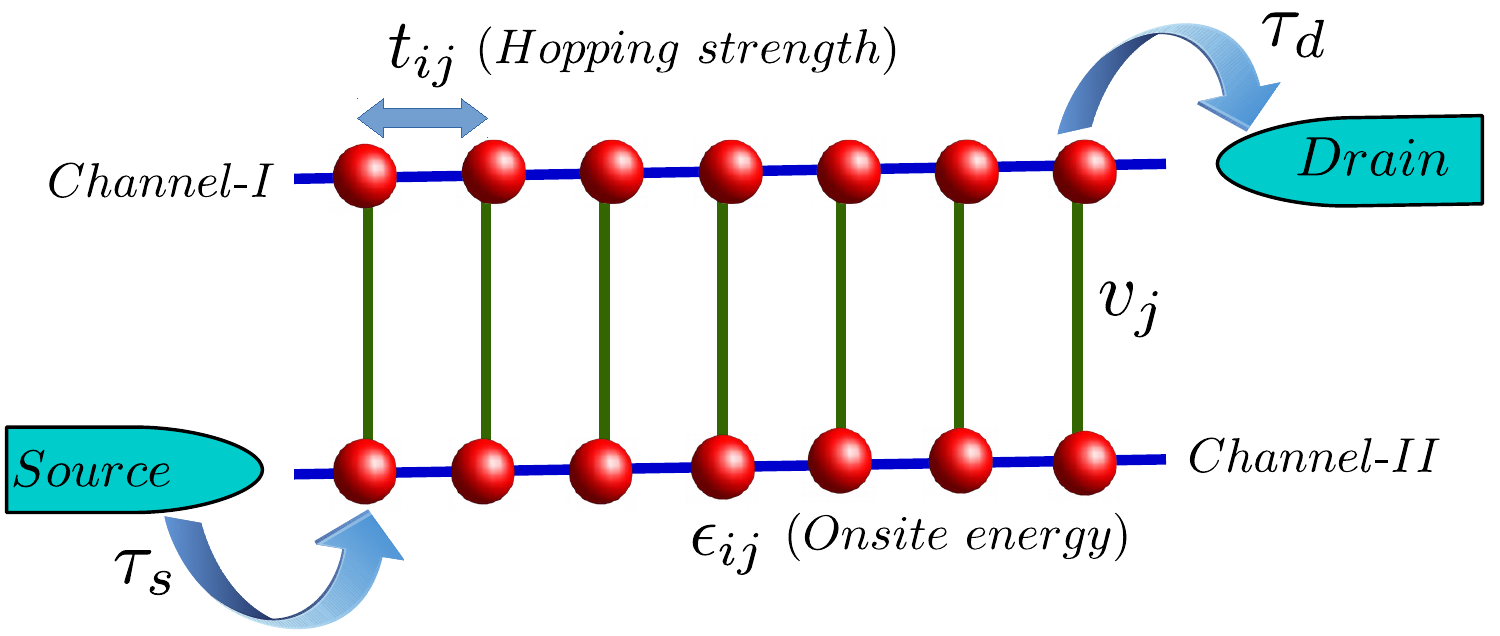}
\caption{(Color online). Junction setup where a double-strand ladder
is coupled to source and drain electrodes.}
\label{mod}
\end{figure}

Let us begin with Fig.~\ref{mod}, where a double-strand ladder is coupled to 
source (S) and drain (D) electrodes. The channel-I and 
channel-II describe the two strands of the ladder those are directly 
connected through vertical bonds (the so-called rungs). We describe the 
junction using a TB framework and within this framework the TB Hamiltonian of 
the full system can be written as a sum $H=H_{ladder}+H_{elec}+H_{tunn}$, 
where $H_{ladder}$ represents the Hamiltonian of the ladder, 
$H_{elec}$ corresponds to the Hamiltonian of the side-attached electrodes, 
and $H_{tunn}$ is associated with the tunnel coupling between the ladder 
and electrodes. 

In absence of e-e interaction, the TB Hamiltonian of the ladder
under NNH approximation reads as
\noindent 
\begin{eqnarray}
H_{ladder} &=& \sum\limits_{i=I,II}\sum_{j}\epsilon_{ij} c^{\dagger}_{ij} 
	c_{ij} + \sum_{i=I,II}\sum_j t_{ij} c^{\dagger}_{ij}
	c_{i,j+1} \nonumber \\
	   & & + \sum_j v_j c^{\dagger}_{I,j} c_{II,j} + h.c.
\label{hamil}
\end{eqnarray}
Here, the index $i$ refers to the strands and $j$ represents the sites in 
each strand. The ladder is parameterized with on-site energy
$\epsilon_{ij}$, intra-strand NNH integral $t_{ij}$ and the inter-strand NNH
integral $v_j$. $c_{ij}^\dagger$ ($c_{ij}$) is the Fermionic creation 
(annihilation) operator at site ($i,j$). In presence of non-zero bias, 
a finite bias drop takes place along the ladder and the actual variation 
of this potential profile is very hard to execute since it involves a 
complete many-body problem. But, we can introduce this effect 
`phenomenologically' into our Hamiltonian, as already proposed earlier
in other works~\cite{dna6,quasi-recti,el1,el2,el3}, considering different 
kinds of potential variation. The most common form of it is the linear 
dependence, and in our work we also consider this variation, though some 
other forms are also taken into account associated with electron screening 
in different materials. With these different choices only the quantitative 
behavior changes slightly, whereas the qualitative features remain unchanged. 
Now, in presence of a finite bias drop along the ladder its site energy
becomes $\epsilon_{ij} = \epsilon_{ij}^0 + \epsilon_{ij}^V$, where 
$\epsilon_{ij}^0$ represents the site energy in absence of bias voltage $V$, 
while $\epsilon_{ij}^V$ is the voltage dependent term. Assuming a linear 
bias drop, the voltage dependent site energy becomes 
$\epsilon_{ij}^V=V/2 - j V/(N+1)$, where $N$ represents the total number 
of rungs in the ladder.

To describe the electronic transport and rectification operation, we connect
the ladder with two reflectionless 1D electrodes. The Hamiltonian of the 
electrodes, parameterized with on-site energy $\epsilon_0$ and NNH integrals 
$t_0$ looks like
\begin{equation}
H_{elec} = \epsilon_{0}\sum_i b^{\dagger}_{iS(D)} b_{iS(D)} + 
t_0 \sum_i b^{\dagger}_{iS(D)} b_{i+1 S(D)} + h.c.
\label{hamil1}
\end{equation}
where $b_{i S(D)}^\dagger$ ($b_{i S(D)}$) is the creation (annihilation)
operator in the electrodes. 

The last part of the Hamiltonian of the full system i.e., the TB Hamiltonian
associated with the tunnel coupling can be expressed as
\begin{equation}
H_{tunn} = \tau_s b_{-1 S}^\dagger c_{I1} + \tau_d c_{II,2N}^\dagger b_{1D} 
+ h.c.
\label{hamil2}
\end{equation} 
where $\tau_s$ and $\tau_d$ are the coupling strengths of the source and 
drain electrodes with the ladder, respectively. In numbering the atomic 
sites of a $N$-rung ladder, we use site number $1$ where the source is 
coupled, while the end site number is $2N$ where the drain is connected.

\subsection{Interacting model}  

The interacting TB Hamiltonian for the ladder neglecting the spin degrees of 
freedom looks like
\noindent
\begin{eqnarray}
H_{ladder} &=& \sum\limits_{i=I,II}\sum_{j}\epsilon_{ij} c^{\dagger}_{ij} 
	c_{ij} + \sum_{i=I,II}\sum_j t_{ij} c^{\dagger}_{ij}
	c_{i,j+1} \nonumber \\
	   & & + \sum_j v_j c^{\dagger}_{I,j} c_{II,j} + h.c + u \sum\limits_{i=I,II}\sum_{j}
	   n_{ij}n_{i,j+1} \nonumber \\
	   & & + u_1 \sum_{j} n_{Ij}n_{II j}
\label{hamilinter}
\end{eqnarray}
where $u$ and $u_1$ are the nearest-neighbor intra-strand and 
inter-strand Coulomb repulsion energy strengths, respectively, and $n_{i,j}$ 
is the number operator. For this open quantum system, we 
treat the interaction term within the Hartree-Fock MF approximation, where 
the occupation number at any site is calculated using self-consistent 
procedure~\cite{inter1,inter2,inter3,inter4}. Within this scheme, the 
interacting system is effectively treated as non-interacting one, and the 
HF Hamiltonian becomes
\noindent
\begin{eqnarray}
H_{ladder}^{HF} &=& \sum\limits_{i=I,II}\sum_{j}\epsilon_{ij}^{\prime} c^{\dagger}_{ij} 
	c_{ij} + \sum_{i=I,II}\sum_j t_{ij}^{\prime} c^{\dagger}_{ij}
	c_{i,j+1} \nonumber \\
	   & & + \sum_j v_j^{\prime} c^{\dagger}_{I,j} c_{II,j} + h.c.
\label{hamilmf}
\end{eqnarray}
Here the modified on-site energies are related to the Hartree 
term and it gets the form $\epsilon_{Ij}^{\prime}=\epsilon_{Ij} + 
u (\langle n_{I,j+1}\rangle +\langle n_{I,j-1}\rangle)+ u_1 n_{II,j}$. 
Whereas, the modified hopping terms are related to the Fock exchange terms 
and they are: $t_{ij}^{\prime}=t_{ij} - u \langle 
c_{i,j+1}^{\dagger}c_{ij}\rangle$ and $v_j^{\prime}=v_j - 
u_1 \langle c_{II j}^{\dagger} c_{I j}\rangle$. 
The MF quantities $\langle n_{ij}\rangle$, 
$\langle c_{i,j+1}^{\dagger}c_{ij}\rangle$ and 
$\langle c_{II j}^{\dagger} c_{I j}\rangle$ are determined self-consistently 
using the NEGF formalism~\cite{inter3}.

\subsection{Theoretical framework for the calculations}

To characterize transport properties and rectification operations, the 
first quantity that we need to calculate is the two-terminal transmission 
probability $T(E)$ ($E$ being the energy of injected particles) across 
the junction. We evaluate it using the NEGF formalism~\cite{green1,
green2,green3,green4,green5}, where the effects of side-attached 
electrodes are incorporated through self-energy corrections. The 
effective Green's function of the non-interacting ladder is written as:
\begin{equation}
G_{ladder}^r = \left[(E+i\eta)I-H_{ladder}-\Sigma_{S}^r - 
\Sigma_{D}^r\right]^{-1}
\label{greens1}
\end{equation}
where $\eta \rightarrow 0^+$, and $\Sigma_S^r$ and $\Sigma_D^r$ are the 
retarded self-energy matrices due to S and D, respectively. From these 
self-energies, we calculate the coupling matrices $\Gamma_S$ and $\Gamma_D$ 
from the expression $\Gamma_{S(D)}=-2\, \mbox{Im}\left[\Sigma_{S(D)}^r\right]$.
Using $\Gamma_S$ and $\Gamma_D$ we compute two-terminal transmission 
probability following the Fisher-Lee relation\cite{green1}, and, it is 
written in the form
\begin{equation}
T = \mbox{Tr} \left[\Gamma_S G_{ladder}^r \Gamma_{D} G_{ladder}^a \right]
\label{greens3}
\end{equation}
where $G_{ladder}^a = (G_{ladder}^r)^{\dagger}$.

In presence of e-e interacting, the Green's function gets 
modified by $G_{ladder}^{HF,r}$ and it becomes,
\begin{equation}
\mathcal{G} = G_{ladder}^{HF,r} = \left[(E+i\eta)I - H_{ladder}^{HF} -
\Sigma_{S}^r-\Sigma_{D}^r\right]^{-1}
\label{greens22}
\end{equation}
$\mathcal{G}$ is a $2N\times2N$ matrix and it is written in $(i,j)$ basis. 
Here $i$ has two values $I$ and $II$ and $j$ runs from $1$ to $N$.
We calculate the different MF quantities $\langle n_{ij}\rangle$, 
$\langle c_{i,j+1}^{\dagger}c_{ij}\rangle$ and 
$\langle c_{II j}^{\dagger} c_{I j}\rangle$ 
from the integration $(1/2\pi)\int M_{pq}\,dE$ where $M_{pq}$ is the
$pq$ (associated with the indices $i$ and $j$) element of the matrix 
$M=\mathcal{G}^{\dagger}\left(\Gamma_S f_S + \Gamma_D f_D \right) \mathcal{G}$.
$f_S$ and $f_D$ are the Fermi distribution functions for S and D, 
respectively. Using
self-consistent procedure, we get the converged $G_{ladder}^{HF}$.
For this interacting system, we calculate the transmission probability 
$T(E)$, similar to the non-interacting one through the expression
\begin{equation}
T=\mbox{Tr} \left[\Gamma_S G_{ladder}^{HF,r} \Gamma_{D} 
G_{ladder}^{HF,a} \right]
\label{greens4}
\end{equation}

In order to inspect the nature of energy bands under different biased 
conditions which is extremely crucial for analyzing rectification mechanism 
we need to compute ADOS, and we perform it using the 
relation~\cite{green2,green3}
\begin{equation}
\rho(E) = - \frac{1}{2N\pi} \mbox{Im} \left[\mbox{Tr} 
\left(G_{ladder}^r\right) \right]
\label{greens44}
\end{equation}
where $2N$ gives the total number of lattice sites in a $N$-rung ladder. 
For the interacting case, $G_{ladder}^r$ will be replaced by 
$G_{ladder}^{HF,r}$.

Finally, we compute transport current across the junction through the 
expression~\cite{green2,green3}
\begin{equation}
I_T = \int dE\, \left[f_S (E) - f_D (E)\right] \,T(E)
\end{equation}

Calculating the currents for two different polarities of external bias we 
eventually compute the rectification ratio ($RR$), and it is defined as
\begin{equation}
RR=\frac{|I_T (+V)| - |I_T (-V)|}{|I_T (+V)| + |I_T (-V)|}
\label{rreq}
\end{equation}
$RR=0$ corresponds to no rectification, whereas $RR=1$ or $RR=-1$ represents 
$100\%$ rectification, depending on which current is fully 
suppressed in a particular bias polarity. 

\section{Numerical Results and Discussion}

In what follows we present our results. We measure all the energies in unit
of electron-volt (eV). Throughout the analysis we choose site energies
of the electrodes $\epsilon_0=0$ and NNH integrals $t_0=3$, unless specified
otherwise, and present $I_T$-$V$ curves in log-linear plot, for better
viewing.

\subsection{DNA molecule}

\begin{figure}[ht]
{\centering \resizebox*{8.5cm}{3.5cm}{\includegraphics{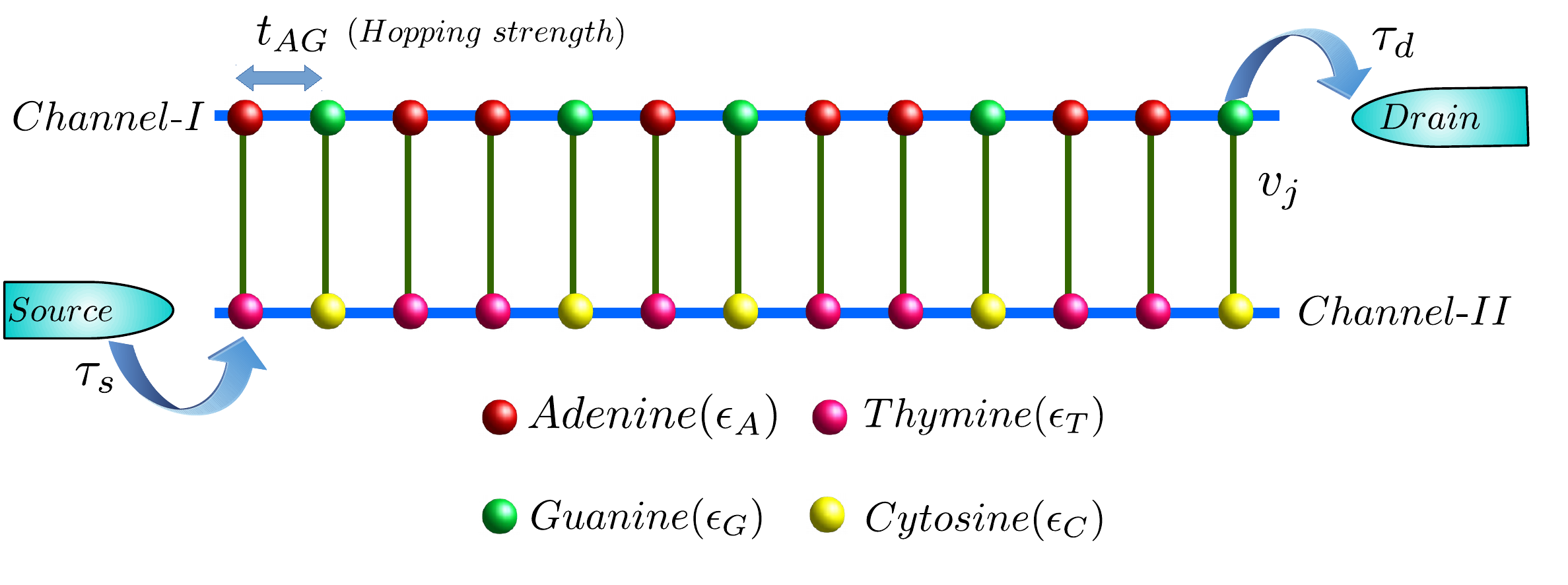}}\par}
\caption{(Color online). dsDNA model represented as two strand ladder with
actual four bases like Adenine, Guanine, Thymine and Cytosine, is coupled
to source and drain.}
\label{fdna}
\end{figure}

First, we want to look into the behavior of current rectification 
for an artificial dsDNA molecule. It is modeled by a two stranded tight-binding 
ladder (see Fig.~\ref{fdna}) considering the four nitrogenous bases, Adenine (A), 
Guanine (G), Thymine (T) and Cytosine (C). We 
choose the site energies of these bases as 
$\epsilon_{G}=8.3$, $\epsilon_{A}=8.5$, $\epsilon_{T}=9.0$, and
$\epsilon_{C}=8.9$. The horizontal hopping strengths between the same bases 
are $t_{GG}=0.11$, $t_{AA}=0.22$, $t_{CC}=-0.05$, and $t_{TT}=-0.14$, while
the horizontal hopping strengths between the different bases can be calculated 
using $t_{AG}=(t_A + t_G)/2$. The vertical hopping strength is set at
$v_j=-0.3$~\cite{dna5}. 

Thousands of various sequences can be composed from these 
bases A, T, G and C. While natural DNA molecules can be extracted from the cells
of all living organisms, artificial DNA molecules could be fabricated in any 
desired sequence. The natural dsDNA is extracted from the sequence of human 
chromosome 22 (chr22). Human chromosome based sequences are hc1, hc2, hc3 
and the sequences consist of the mixture of all the four bases. The man-made 
dsDNA is taken in the form of a random sequence as well as in substitutional
form like nickel mean, copper mean, triadic cantor, Fibonacci, etc~\cite{dna5}. 
All these substitutional sequences are constructed following an inflation rule. 
For instance, starting from the base A, Fibonacci sequence can be formed by 
the rule A$\rightarrow$AG and G$\rightarrow$A. Thus, the substitutional 
sequences for a strand are the mixture of two base pairs. Knowing the 
sequence for one single-strand, one can construct the sequence for the 
other strand satisfying the Watson-Crick (WC) base-pairing rules: G pairs 
with C, and A pairs with T~\cite{dna5}. The substitutional sequences form 
quasi-periodic potential. In Ref.~\cite{quasi-recti}, we have seen 
quasi-periodicity plays an important role behind large rectification 
in a 1D chain. Therefore, to construct the artificial dsDNA molecule, 
we arrange the bases A and G in the upper strand (i.e., channel-I) 
following a Fibonacci sequence (quasi-periodic one), and 
the other strand is automatically designed with the bases T and C
(which is also a quasi-periodic sequence) following the WC base-pairing 
rules~\cite{dna5}. 
To explore the rectification operation we connect the dsDNA molecule with the
two electrodes (see Fig.~\ref{fdna}), those are parametrized with on-site 
energy $\epsilon_0=0$ and NNH integrals $t_0=5$. 
\begin{figure}[ht]
{\centering \resizebox*{7cm}{7cm}{\includegraphics{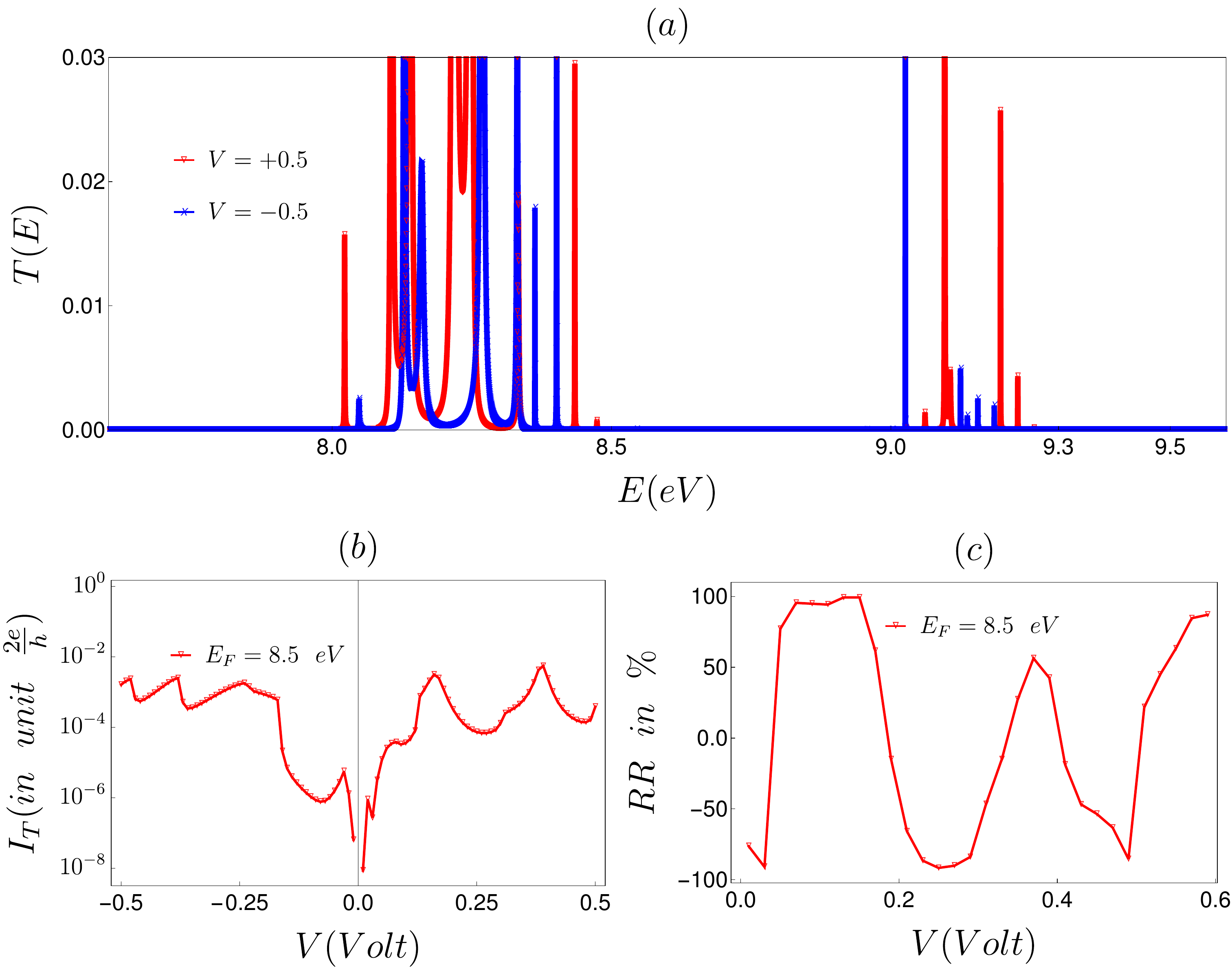}}\par}
\caption{(Color online). Results of the Fibonacci dsDNA molecule having 
$13$ rungs where (a), (b) and (c) correspond to the transmission probability, 
junction current and the variation of rectification ratio.}
\label{f15}
\end{figure}
The molecule-to-electrode coupling is fixed at $\tau_s=\tau_d=0.7$. 
The results for this molecular junction, computed at zero temperature and
in the absence of e-e interaction, are placed in Fig.~\ref{f15}. Here we 
find distinct nature of transmission spectra under two bias polarities 
(Fig.~\ref{f15}(a)). In reversing the bias, the energy spectrum of the 
bridging system gets changed, as site energies are voltage dependent, 
which results different transmission spectra under two biased conditions. 
As the spectrum is gapped due to the quasi-periodic nature of site energies, 
one can definitely find a suitable Fermi energy $E_F$ about which 
transmission is finite over a small energy window for one bias polarity, 
while it becomes vanishingly small or even zero for the other biased 
condition. This leads to a high degree of current rectification which is 
exactly shown in Figs.~\ref{f15}(b) and (c). At some specific voltage 
regions almost $100\%$ rectification ($RR \sim \pm 100\%$) is obtained, 
which has not been reported so far in literature. 
In their work Guo {\em et al.}~\cite{dna6} have shown 
that the artificially designed dsDNA with $11$ base pairs, where three
different bases are mixed (CGCGAAACGCG), cannot produce strong current
rectification. But, after intercalating two coralyne molecules into the 
dsDNA one, the whole DNA-coralyne complex exhibits much stronger 
rectification. In comparison to dsDNA molecule, DNA-coralyne complex 
shows higher `spatial asymmetry' at the left and right edges where 
electrodes are connected to the system. The striking spatial asymmetry 
induced by coralyne intercalation is the explanation for the strong
rectification in the DNA-coralyne complex. In our model, the sequence 
of dsDNA molecule and the mechanism behind rectification are completely 
different from what is discussed in Ref.~\cite{dna6}. We show that the
artificial dsDNA with Fibonacci type substitutional sequence exhibiting 
strong rectification depending on the gapped and fragmented nature of
energy spectrum, solely depends the quasi-periodic behavior of 
substitutional sequence. For the other substitutional sequences also, 
it is possible to have the perfect current rectification as the physics 
behind rectification remains unchanged. It gives a clear signature that 
the artificial dsDNA molecule with substitutional sequences can be a 
suitable functional element for utilizing rectifying operation at 
nanoscale level. Other than the dsDNA, different bio-molecular systems 
like nucleic acids and most proteins follow the quasi-periodic 
orders~\cite{macia06}. As the quasi-periodicity gives rise to strong 
rectification, the other bio-molecular systems may also be implemented 
for strong rectification.

\subsection{Ladder with AAH potential}
\label{AAH_ladder}

As quasi-periodicity plays an important role in getting high degree of
rectification, which is established from the results of DNA molecule, now
we focus our attention on the current rectification considering an AAH 
ladder. It may provide an additional advantage as the rectification ratio
can be tuned {\em externally} with the AAH phase, which is always helpful
in designing a device.
AAH potential is also a quasi-periodic potential and it has the form 
$\epsilon_{i}=W \cos\left[2\pi b i + \phi_{\beta}\right]$~\cite{aubry1}. 
Here $W$ is the strength of the potential, $\phi_{\beta}$ is the 
phase associated with the AAH modulation, and, $b$ is an irrational number. 
It is important to note that the phase factor 
$\phi_{\beta}$ does not change the behavior of transport, but it changes 
the location of the bands in presence of a linear potential drop.
In our calculations we set $b=(1+\sqrt{5})/2$, the golden mean, which
is most commonly used, though any other irrational number can be taken 
into account and the physics will not be changed at all. Depending on the
strength $W$, the AAH system contains extended, critical and localized 
states. Transport through the extended states are faster than the critical 
states. With increasing system size transport through 
critical states gets decreased, whereas for the extended ones as the 
transport is ballistic it does not depend on the size of the system. 
This is one of key the advantages of the AAH type quasi-periodic 
sequence over the other quasi-periodic sequences like Fibonacci, 
Thue-Morse, etc., as these discrete quasi-periodic sequences only 
posses critical single particle eigenstates~\cite{fragment1}. 

An AAH ladder can in principle be designed by incorporating AAH modulations 
in different parts like on-site energies and/or inter or intra strand 
NNH integrals. Now we will focus on a specific configuration where we set 
$\epsilon_{Ij}^0 = W \cos\left[2\pi b j + \phi_{\beta}\right]$, 
$\epsilon_{II j}^0$=0, $t_{Ij}=t_{IIj}=t$ and $v_j=v, \forall j$.

\subsubsection{Rectification}

\begin{figure}[ht]
{\centering \resizebox*{7.5cm}{7cm}{\includegraphics{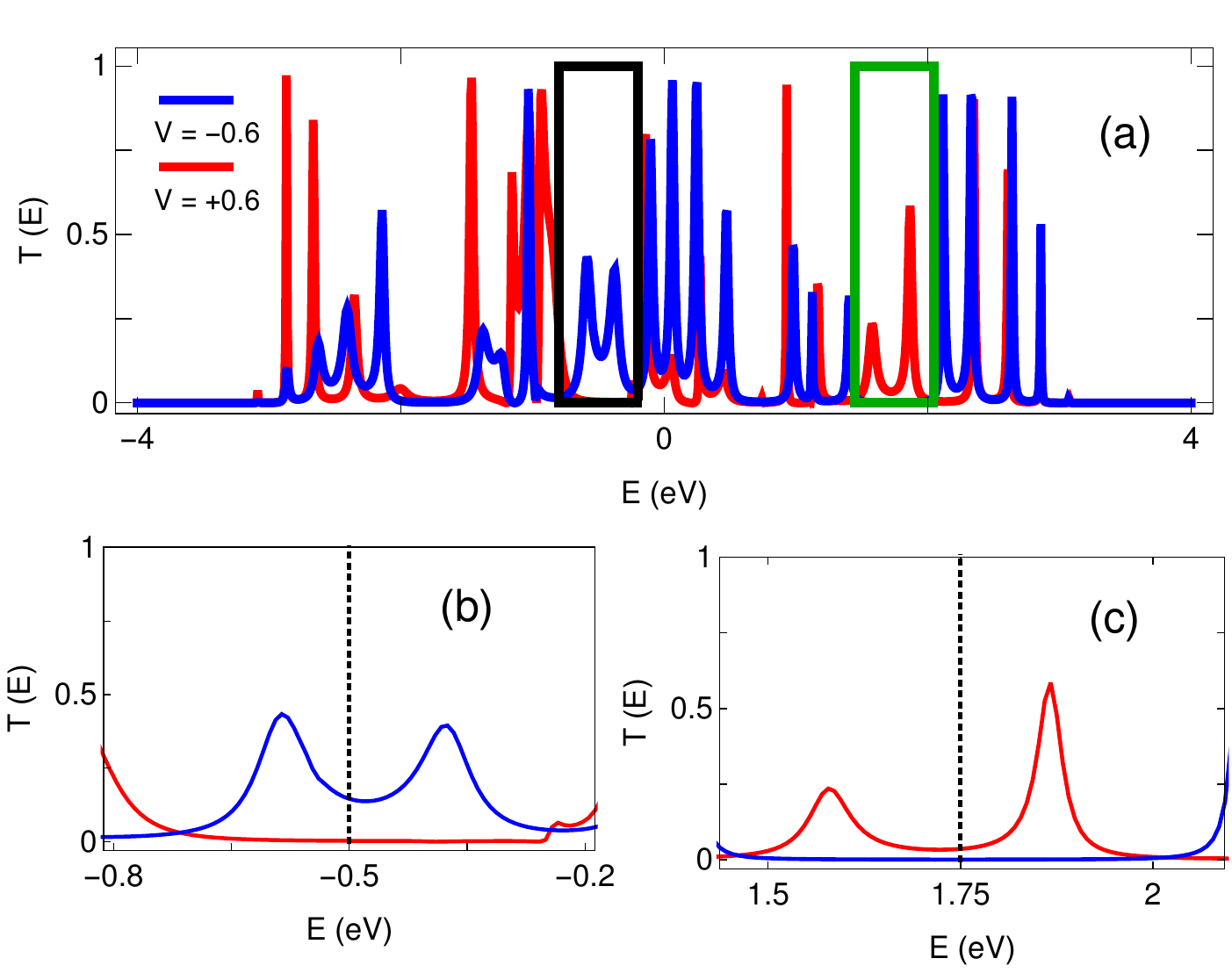}}\par}
\caption{(Color online). (a) Two-terminal transmission 
probability as a function of energy for a $13$-rung ladder under forward and 
reverse bias conditions, where the red and blue lines correspond to $V=0.6\,$V
and $-0.6\,$V, respectively. The spectra given in (b) and (c) represent the 
enlarged versions of the black and green framed regions of (a), respectively,
for clear viewing of different colored curves in these two energy 
zones of our interest. The imaginary dotted lines in (b) and (c) are associated
with the locations of Fermi energy. The other physical parameters taken for 
these calculations are: $t=1$, $v=1$, $W=1$, $\phi_{\beta}=0$, $\tau_s=1$, 
and $\tau_d=0.6$. Here we choose asymmetric conductor-to-electrode coupling.}
\label{f1}
\end{figure}

To show the rectification behavior, in Fig.~\ref{f1}(a) we show the variation
of transmission probabilities of a ladder network at a typical bias voltage
under forward and reverse bias conditions. At a first glance we see that
the transmission spectra are quite different for two distinct polarities
of the external bias, providing
sharp resonant peaks at multiple energies. A careful inspection reveals that
there are some (small) energy windows where finite transmission is obtained
only due to one bias polarity, while the transmission gets almost
suppressed in other bias condition, and this phenomenon will lead to much
higher rectification. To examine this fact, we
selectively choose two small energy zones from Fig.~\ref{f1}(a), marked by the
black and green framed regions, and replot the enlarged versions in 
Figs.~\ref{f1}(b) and (c), respectively. Interestingly we see from these 
spectra (Figs.~\ref{f1}(b) and (c)) that, for the entire energy zones one 
curve (blue or red) dominates, whereas the other one almost vanishes.
\begin{figure}[ht]
{\centering \resizebox*{7.5cm}{6.5cm}{\includegraphics{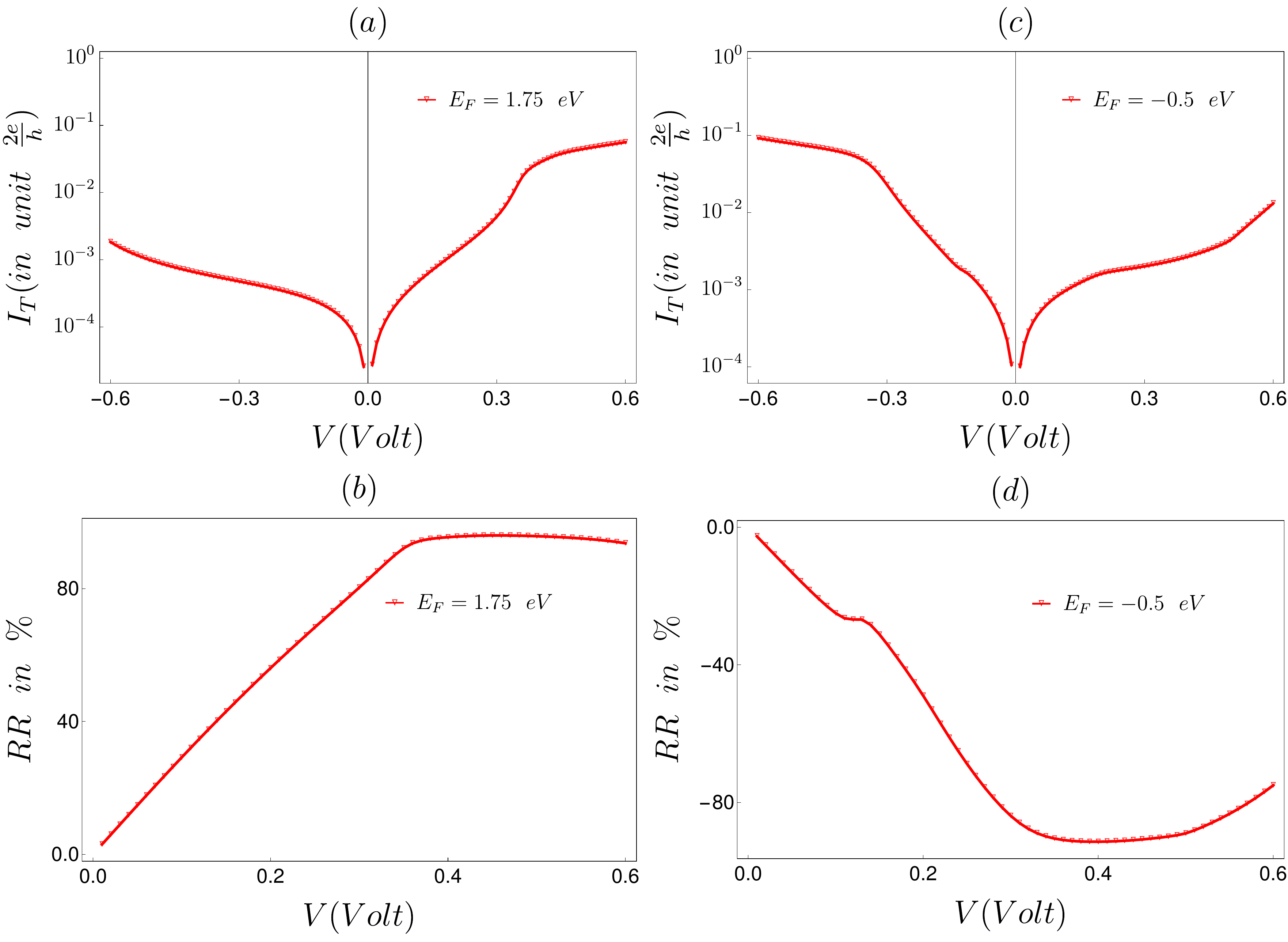}}\par}
\caption{(Color online). $I_T$-$V$ and corresponding
$RR$-$V$ characteristics at two different Fermi energies for the AAH ladder
considering the identical physical parameters as taken in Fig.~\ref{f1}.}
\label{f2}
\end{figure}
\begin{figure}[ht]
{\centering \resizebox*{6cm}{7.5cm}{\includegraphics{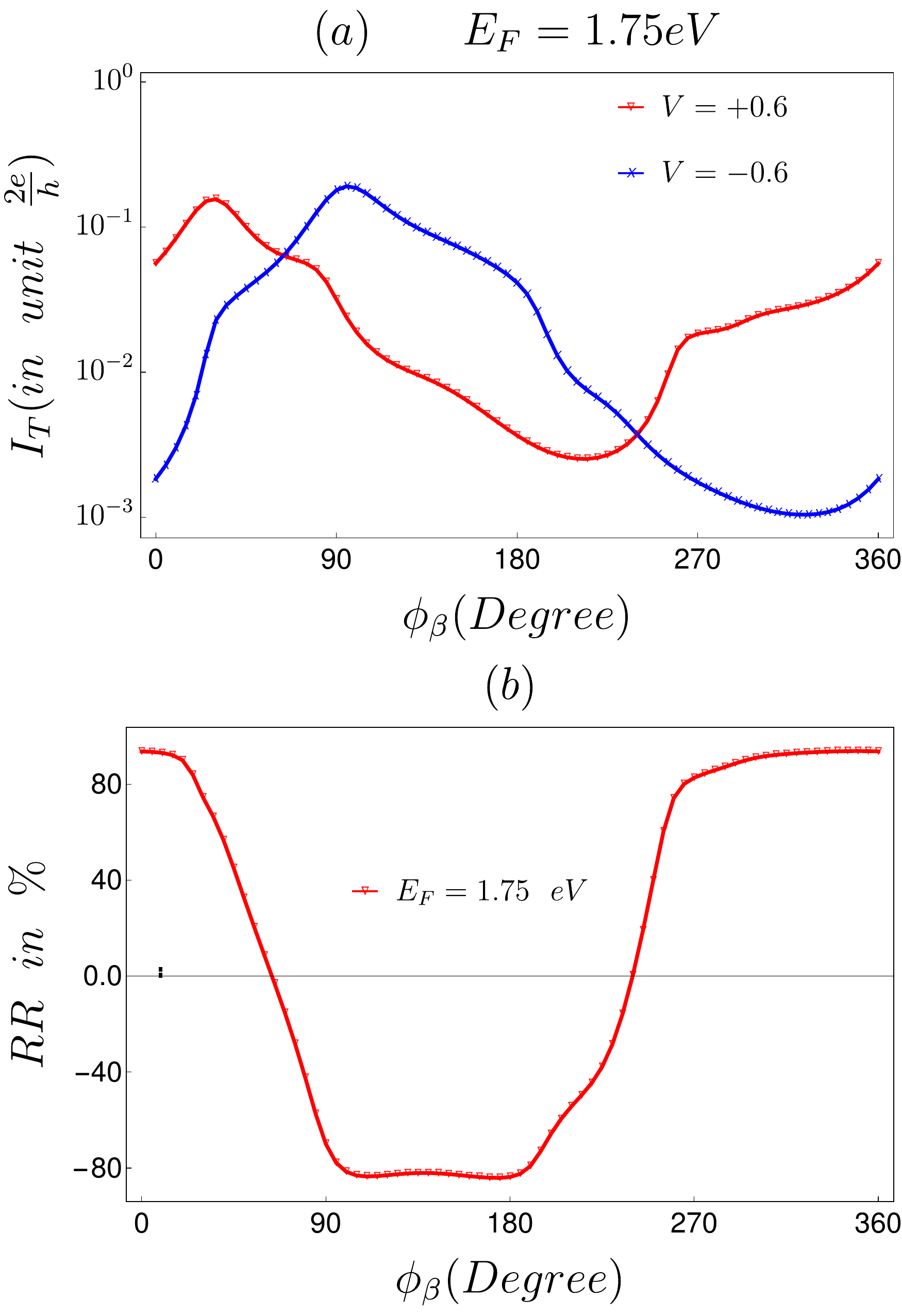}}\par}
\caption{(Color online). Role of AAH phase $\phi_{\beta}$ on transport 
current and rectification ratio. In (a), transport currents are shown for 
two bias polarities, represented by two different colored curves in a 
log-linear plot, and the corresponding $RR$ is given in (b). Here we 
choose $E_F=1.75\,$eV, and all the other physical parameters are kept 
unchanged as taken in Fig.~\ref{f1}.}
\label{f3}
\end{figure}
To inspect the dependence of rectification for other voltages, let us focus
on the spectra given in Fig.~\ref{f2} where we calculate transport current
along with rectification ratio by varying the voltage from zero to $0.6\,$V
under two bias polarities, setting the Fermi energy at two suitable energies.
For $E_F=1.75\,$eV, the transport current in the negative bias condition is 
vanishingly small, whereas much higher current is obtained in the positive
bias condition (see Fig.~\ref{f2}(a)). As a result of this, {\em very high 
degree of $RR$ is achieved, and most remarkably, this feature persists over 
a broad range of voltage bias} (Fig.~\ref{f2}(b)). The phenomenon gets 
almost reversed when we set the Fermi energy at $-0.5\,$eV, as clearly 
noticed from the spectra Figs.~\ref{f2}(c) and \ref{f2}(d). Thus, $E_F$ 
is one of the tuning parameters for regulating $RR$. 
If $\mu_1$ and $\mu_2$ are the chemical potentials 
of the source and drain, respectively, and $V$ is the voltage bias 
maintained by the difference in chemical potentials, then we can write 
$\mu_1=E_F+V/2$ and $\mu_2=E_F-V/2$. In experiments, $E_F$ is changed 
routinely in-situ by controlling external gate voltage~\cite{pr26}.

In Fig.~\ref{f3} we describe the critical role of the AAH phase $\phi_{\beta}$ 
on transport current and rectification ratio. The transport current is highly 
sensitive to the AAH phase, and currents are markedly different for two 
distinct bias polarities (see Fig.~\ref{f3}(a)).
This behavior is clearly reflected in the $RR$-$\phi_{\beta}$ spectrum
(Fig.~\ref{f3}(b)). From Fig.~\ref{f3} we can see that {\em the 
$RR$ achieves a very high value, close to $100\,\%$, and almost complete 
phase reversal ($RR$ reaches to $-100\,\%$) takes place upon the variation 
of $\phi_{\beta}$.} {\em The other important feature is that the high degree 
of rectification persists over a wide range of the phase factor 
$\phi_{\beta}$, that gives a clear hint that one can design a setup where 
$\phi_{\beta}$ can be selected from a reasonable window, and sharp tuning 
is no longer precisely required.} Here we would like to 
note that $\phi_{\beta}$ can be changed by designing different realizations 
of the on-site potential in a same experimental setup (details are obtained 
in the supplementary material of Ref.~\cite{expt0} and other 
works~\cite{expt1,expt2,expt3}. 

\subsubsection{Reason for 100$\%$ rectification}

The large degree of rectification is caused due to the two reasons: 
(i) gapped fragmented spectrum and spatial reflection symmetry breaking 
of quasi-periodic potential, and (ii) voltage dependence of the energy 
levels due to the term $\epsilon_{ij}^V$. Depending on these two factors, 
at zero temperature it is always possible to get a Fermi energy about 
which there is an energy window where finite transmission is present for 
\begin{figure}[ht]
{\centering \resizebox*{7.5cm}{8cm}{\includegraphics{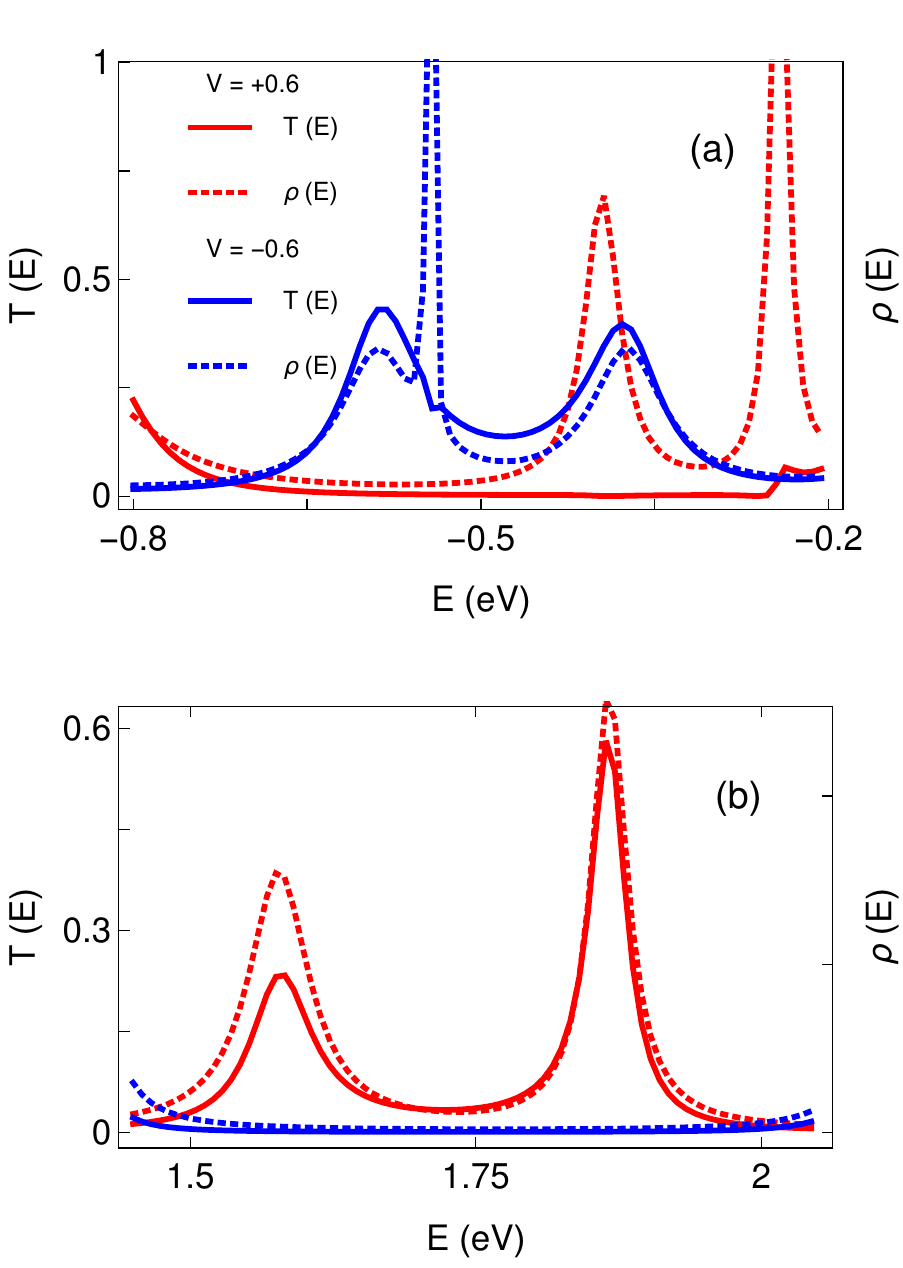}}\par}
\caption{(Color online). Transmission probability $T(E)$
and ADOS $\rho(E)$ for the two energy zones (shown in the upper
and lower panels) as considered in Figs.~\ref{f1}(b) and (c), respectively.
All the physical parameters are same as taken in Fig.~\ref{f1}.}
\label{f1new}
\end{figure}
one biased condition and gapped for the other biased condition. This leads 
to $100\%$ rectification which is clearly reflected from Fig.~\ref{f1} 
and Fig.~\ref{f2}. The other possible reason of getting enhanced 
rectification can be in some cases due to the Wannier-Stark (WS) 
localization~\cite{wannier1,wannier2,wannier3}. In presence of a non-zero 
bias, site energies are (electric) field dependent yielding a non-uniform 
distribution of the site energies which causes particle localization, 
analogous to a conventional disordered lattice, which is known as WS 
localization. The WS localization at finite bias can be seen from the 
spectra given in Fig.~\ref{f1}, and more clearly they can be visualized 
from the results presented in Fig.~\ref{f1new} as here we superimpose 
ADOS along with the transmission function. The red-dotted line is 
markedly different from the blue-dotted one associated with the energy 
shifting at two bias polarities. Where the dotted line (red and/or blue)
vanishes (viz, states are no longer available), the transmission probability 
naturally drops to zero. Whereas, for the energies when the transmission 
function vanishes irrespective of finite density of states we can conclude 
that the states associated with those energies are localized, and the
localization is caused as a results of electric field (WS localization).
{\em Thus, finding of an energy zone (or more zones) where transmission 
probability is finite for one bias polarity and zero (or vanishingly small) 
for other polarity is always expected depending on the above reasons.}

To the best of our knowledge, no one has addressed before this high degree 
of rectification, for such a wide range of bias voltage. Thus it brings a
significant impact in the era of nanotechnology. At finite temperatures, 
the contributions from other energy zones will also appear which may 
reduce the rectification ratio, and we will discuss these issues in the
forthcoming subsections.

To make the analysis more realistic and keeping in mind possible experimental
realizations of the proposed model, we need to incorporate other different
physical factors that may available in practical situations. Below we discuss
these effects one by one as follows.

\subsubsection{Critical roles of different physical factors on 
rectification}

Here we want to address the effects system size, temperature and 
electron-electron interaction on current rectification, and to test how the
above presented results are modified with these factors.

\vskip 0.2cm
\noindent
\underline{Effect of system size}:
As the energy spectrum of the AAH system is always gapped irrespective
\begin{figure}[ht]
{\centering \resizebox*{6cm}{7.5cm}{\includegraphics{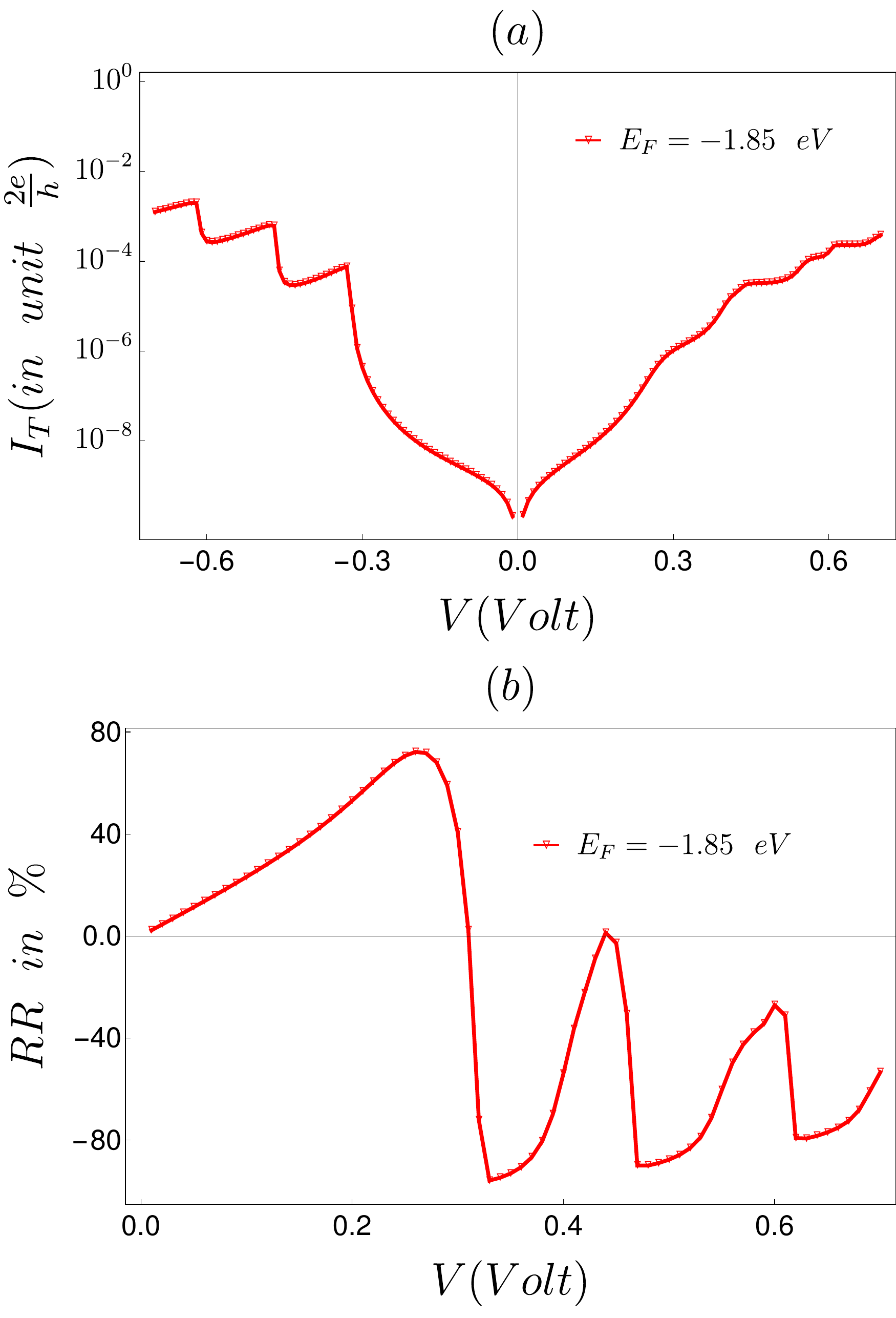}}\par}
\caption{(Color online). $I_T$ and corresponding $RR$
as a function of voltage bias for a bigger AAH ladder with $55$ rungs
($110$ sites). Here we choose $E_F=-1.85\,$eV. The other parameters are
same as taken in Fig.~\ref{f1}.}
\label{faa}
\end{figure}
of the system size, we can still expect a very high degree of current 
rectification even for much bigger AAH ladder. To substantiate this 
fact in Fig.~\ref{faa} we plot the transport current and corresponding 
rectification ratio considering a bigger ladder with $55$ rungs
(110 sites). A high degree of rectification is clearly noticed for
this ladder, and thus we can suggest that the basic features of
rectification are not restricted within a specific system size, rather
we can vary it in wide range which gives a suitable hint for testing
the results in a laboratory setup.

\begin{figure}[ht]
{\centering \resizebox*{7.5cm}{4.5cm}{\includegraphics{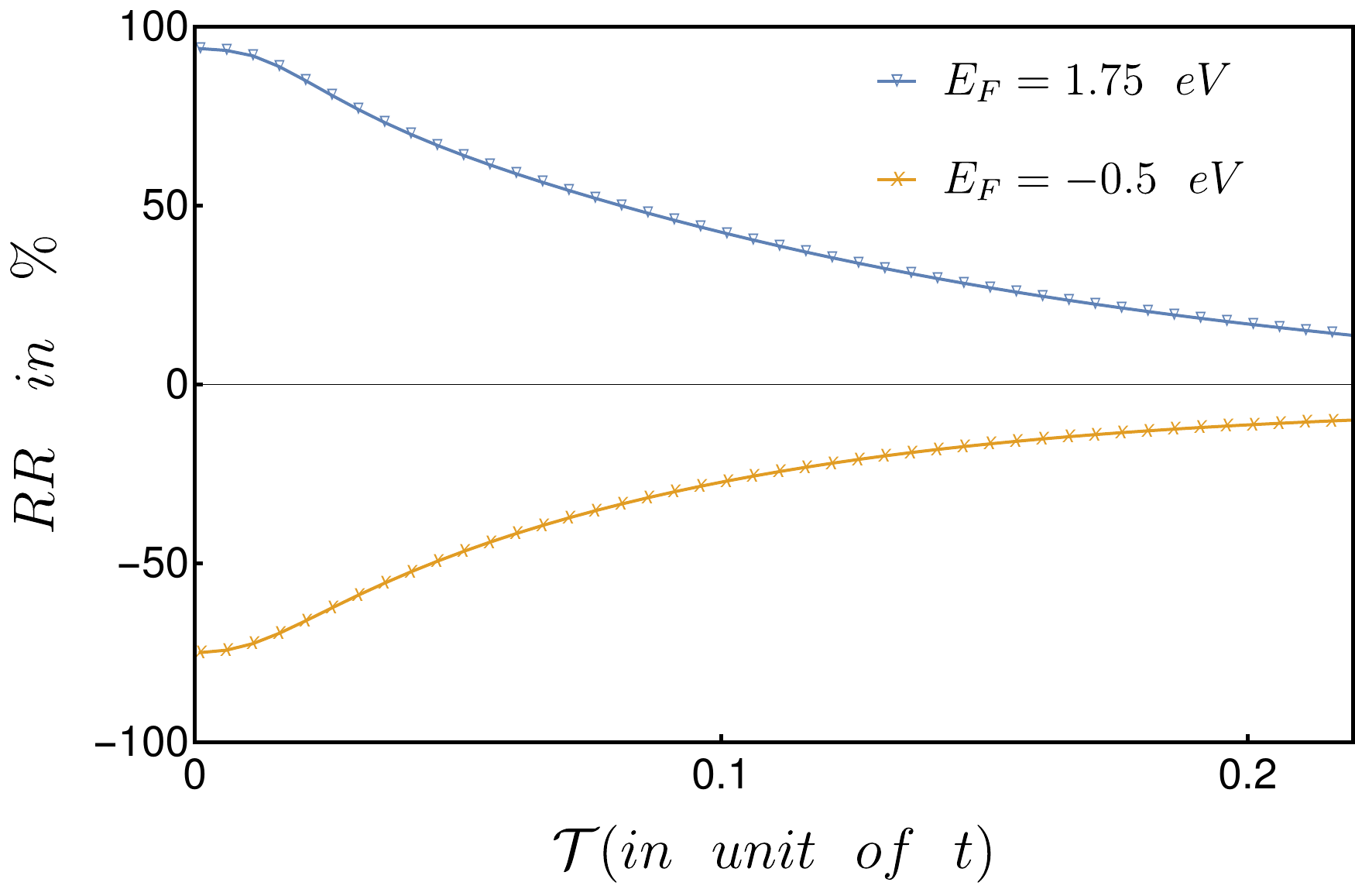}}\par}
\caption{(Color online). Temperature dependence of rectification ratio.
The Fermi energies and other parameters are chosen as in Fig.~\ref{f1}.}
\label{f13}
\end{figure}

\vskip 0.2cm
\noindent
\underline{Effect of temperature}: The results studied so far are 
worked out at absolute zero temperature. Now, we include the effect of 
temperature and study the characteristic features of rectification. The 
results are shown in Fig.~\ref{f13} where we plot the $RR$ as a function 
of system temperature for two different Fermi energies. The $RR$ gets 
decreased with the enhancement of temperature. Though it ($RR$) gets 
reduced, still sufficiently large $RR$ persists over a
moderate range of temperature. This reduction of the degree of rectification 
can be implemented from the argument that at finite temperature we need to
incorporate all the energy levels in the allowed energy window, unlike the
case of zero temperature which excludes the possibility of integrating 
transmission function in a specific energy window where finite transmission 
is obtained for one bias polarity. Thus, it is no longer possible to get 
absolutely $100\%$ rectification since finite contributions from 
both the bias polarities are involved.

\begin{figure}[ht]
{\centering \resizebox*{6.5cm}{11cm}
{\includegraphics{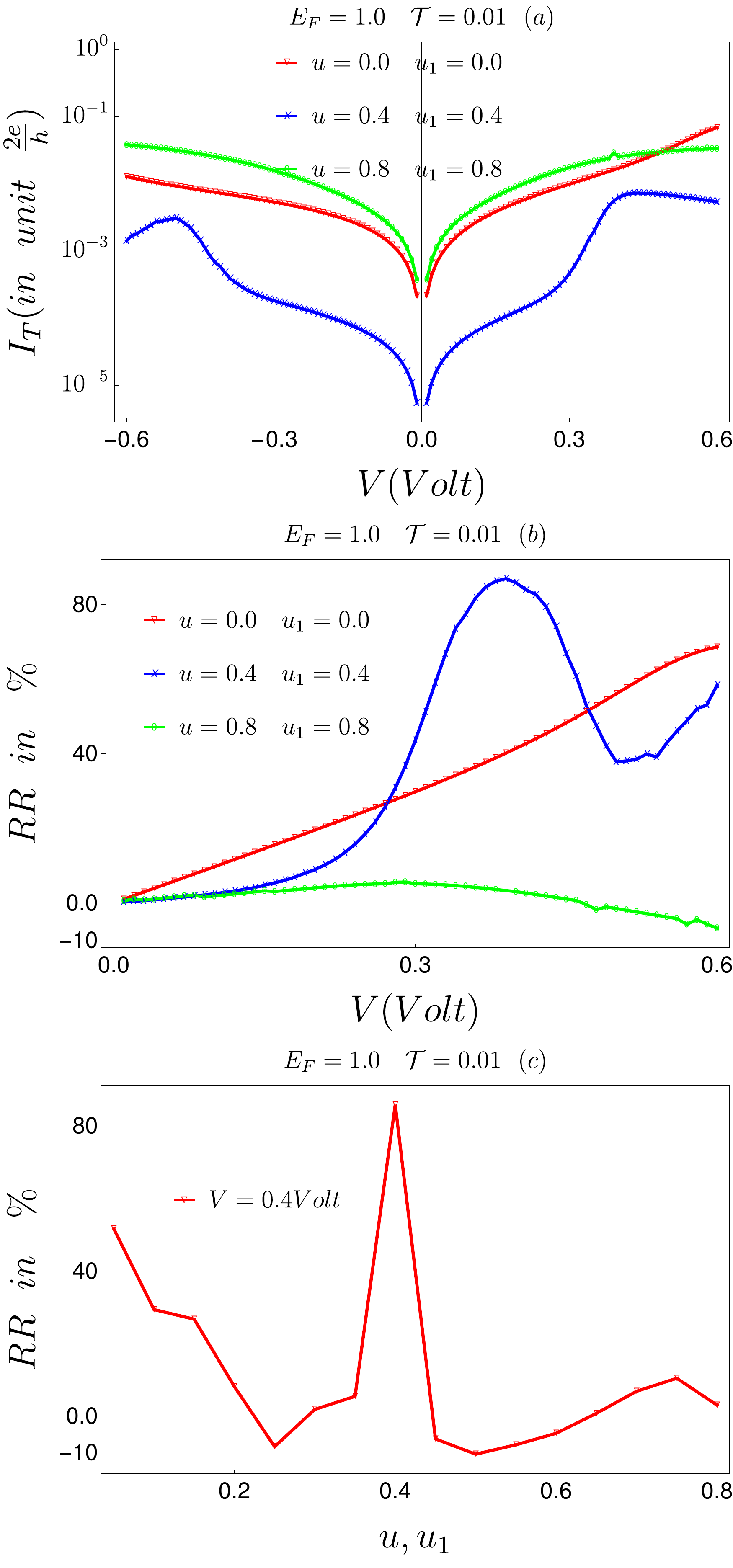}}\par}
\caption{(Color online). Transport current $I_T$ in (a) and the corresponding
$RR$ in (b) as a function of bias voltage for some typical strengths of $u$
and $u_1$. In (c), the variation of $RR$ with $u$ ($=u_1$) is shown when the
bias voltage is fixed at $V=0.4\,$V. The results are computed for a AAH ladder
with five rungs setting the temperature at $0.01$ (in unit of $v$).}
\label{finter}
\end{figure}

\vskip 0.2cm
\noindent
\underline{Effect of electron-electron interaction}: Finally, we include 
electron-electron (e-e) interaction into the system and study its effect on
rectification. We incorporate this effect at the MF level. It is well-known 
that the study of an open interacting system is a state-of-the-art research 
problem at formalism level. Even in the MF limit, one needs to do numerical 
integrations of the NEGF to evaluate the MF quantities, which is an extremely 
challenging task especially for large size systems exhibiting fragmented 
energy spectrum. Due to this fact here we restrict ourselves to the small 
system size. We present the results in Fig.~\ref{finter} for an AAH ladder 
considering five rungs at some typical values of interaction strengths. 
The key feature is that, at some typical values of e-e interaction, there 
is a finite possibility to achieve a very high degree of rectification 
compared to the interaction free system, which is clearly visible from 
Fig.~\ref{finter}(b). It immediately raises an obvious curiosity that how
the degree of rectification modifies with the gradual change of e-e 
interaction strength. To scrutinize it in more detail, in Fig.~\ref{finter}(c)
we plot the rectification ratio by continuously varying the strength of
$u$ ($=u_1$) in a wide range. From the spectrum, it is 
clear that even in presence of electron-electron interaction $(u, u_1)$, 
it is possible to have finite current rectification.
Thus, at the MF level we can see that e-e interaction does not suppress the 
rectification action, rather better performance could be expected for some
typical values of $u$ ($u_1$). To estimate these specific values, more 
detailed analysis should be required and we leave this job for our future 
works.

Here we would like to note that, as already stated above, the computations of
mean-field quantities are heavy for large interacting systems, and thus,
we present the results considering a small size ladder. But, the essential 
physics will remain same for a reasonably large system as well, since the 
degree of rectification is essentially depends of the ratio of currents
under two bias polarities. When the system size is very small, comparable
to a quantum dot (QD), one may expect the effect of Coulomb 
blockade~\cite{cb1,cb2,cb3}, especially for the situation when longer-range 
interactions are present and the system temperature is very low. In presence 
of this effect, although
the electrical conductance, and thus, the junction current gets reduced,
we do not expect any non-trivial feature in the degree of rectification
as it involves the ratio of two currents which are equally affected by 
Coulomb repulsion. The same argument is also valid even if we consider 
the precise role of Coulomb drag~\cite{cd1,cd2,cd3} in our analysis. 
This dragging mechanism 
is generally expected when the two layers or channels are sufficiently 
close to each other, typically within the range of $1$-$2$ nanometers. 
So, in one situation we may think that if we design an AAH ladder where 
two strands are well separated, we can safely ignore the effect of Coulomb 
drag. On the other hand, for the situation when both the two strands are 
close enough, Coulomb drag effect will be an important one which will modify
the currents. But as the ratio of currents under two biased conditions are
taken into account we expect similar kind of rectification operation.

In the sub-section~\ref{AAH_ladder}, we consider a 
double-stranded ladder with AAH modulation in Channel-I. For this ladder we
establish a complete rectification operation depending on the choice of 
Fermi energy $E_F$ at zero temperature, solely due to the quasi-periodic 
nature of the potential. Thus, it is expected that employing quasi-periodicity 
in any segment of the ladder leads to complete current rectification at 
zero temperature. To check that we include quasi-periodicity in different 
segments of ladder and the results are critically analyzed in Appendix~\ref{aa} 
and Appendix~\ref{bb}. Similar kind of complete current rectification is
noticed, depending on the choices of $E_F$. We also show that changing 
the arbitrary phase term in the continuous quasi-periodic potential, the 
magnitude and direction of the rectification ratio can be varied significantly. 
In sub-section $a$ of Appendix~\ref{aa}, we consider the AAH modulation only 
in the inter-strand hopping strengths. The phase of the AAH potential is
referred as $\phi_{\lambda}$ when we consider the modulation in hopping
integrals. We take into account the AAH modulations in on-site energies 
(sub-section $b$ of Appendix~\ref{aa}), and, in intra-strand hopping
integrals (sub-section $c$ of Appendix~\ref{aa}) of both the channels. 
In the previous cases, we consider AAH modulation either in the
on-site potentials or in the hoppings. Quasi-periocity can also be
included both in the on-site potential and hopping strengths, and the
combined effect of both the two phases, $\phi_{\beta}$ and $\phi_{\lambda}$,
leads to interesting behavior in particle current rectification. We explore it
in detail in Appendix~\ref{bb}.

\section{Summary and outlook}

In this work, we have studied particle current rectification 
for a tight-binding double-stranded ladder using NEGF formalism with discrete 
and continuous quasi-periodic potentials. We have started our analysis with 
an artificial dsDNA molecule which has Fibonacci type of substitutional sequence. 
This is a discrete quasi-periodic potential. The quasi-periodicity yields 
gapped and fragmented spectrum, which, in turn, leads to complete rectification 
on tuning the gate voltage at zero temperature for this artificial dsDNA. 
This is a new mechanism of getting rectification, and it greatly differs from 
the previous works (such as~\cite{dna6}) available in literature. Other 
biomolecular systems like nucleic acids, most of the proteins are also 
made up of quasi-periodic sequences. Our work thus suggests that they 
can also be used as functional elements to get suitable rectifications.

In the next part we have considered a double-stranded ladder 
with the continuous quasi-periodic AAH potential instead of discrete Fibonacci 
sequence. This system has two added advantages. Discrete quasi-periodic 
systems exhibit `critical' (intermediate between extended and localized) wave 
functions. Transport through such wave-functions decreases with increase 
in system size. A system with a weak AAH potential provides completely 
extended wave functions, while also having gapped and fragmented spectrum 
those are required for current rectification. The transport through such 
extended wave functions is ballistic and does not depend on the system size. 
The AAH potential has a phase factor, which does not change the transport 
behavior, but changes the location of the bands in presence of a linear 
potential drop. Thus, this phase factor in the potential can serve as an 
additional tuning parameter for the regulation of current rectification, 
unlike the discrete quasi-periodic sequences where Fermi energy acts as 
the only controlling parameter. The AAH phase can be changed by design. 
As, the rectification behavior is governed by the 
quasi-periodicity, we have found complete rectification introducing 
quasi-periodicity at different segments of the ladder network. 
Non-interacting systems with AAH quasi-periodic potential have been 
realized experimentally~\cite{expt0,expt1,expt2,expt3}. Our results 
point to the possibility of large tunable rectification in such experiments.

Apart from complete rectification, another very important 
salient feature we have explored is that the direction of rectification 
can be changed just by controlling the gate voltage in such systems with 
discrete or continuous quasi-periodic potentials. Complete rectification 
due to quasi-periodicity occurs at zero temperature. But, in real experimental 
setups the effect of temperature is unavoidable. Considering finite 
temperature, we have shown that the rectification ratio gets decreased, but 
still it is possible to get a reasonable amount of rectification for a 
moderate range of temperature. Finally, we have considered the effect of 
nearest-neighbor repulsive interaction for spinless fermions at the 
mean-field level. The mean-field results suggest that the interaction 
leads to a reduction in transport, but still a finite rectification is 
possible depending on the interaction strength.

\section{Acknowledgments}

MS would like to thank Archak Purkayastha for useful discussions and 
would like to acknowledge University Grants Commission (UGC) of India 
for her research fellowship.
SKM respectfully acknowledges the financial support of the Science and 
Engineering Research Board, Department of Science and Technology, 
Government of India (Project File Number: EMR/2017/000504).

\appendix

\section{Different other configurations of AAH ladder}
\label{aa}

\subsubsection{AAH modulation in inter-strand hopping}
\label{AAH_inter_strand}

In this configuration, both the two strands are perfect (viz, 
$\epsilon_{Ij}^0=0$ and $\epsilon_{IIj}^0=0$) and the intra-strand 
hopping integrals are uniform i.e., $t_{Ij}=t_{IIj}=t \, \forall \,j$. 
For this network we introduce modulation in the inter-strand hopping 
integrals in the form: $v_j = W_1 \cos\left[2\pi b j + 
\phi_{\lambda}\right]$, where $W_1$ is the modulation strength and 
$\phi_{\lambda}$ is the phase factor that can be tuned 
selectively~\cite{aubry5}.
\begin{figure}[ht]
{\centering \resizebox*{7.5cm}{7cm}{\includegraphics{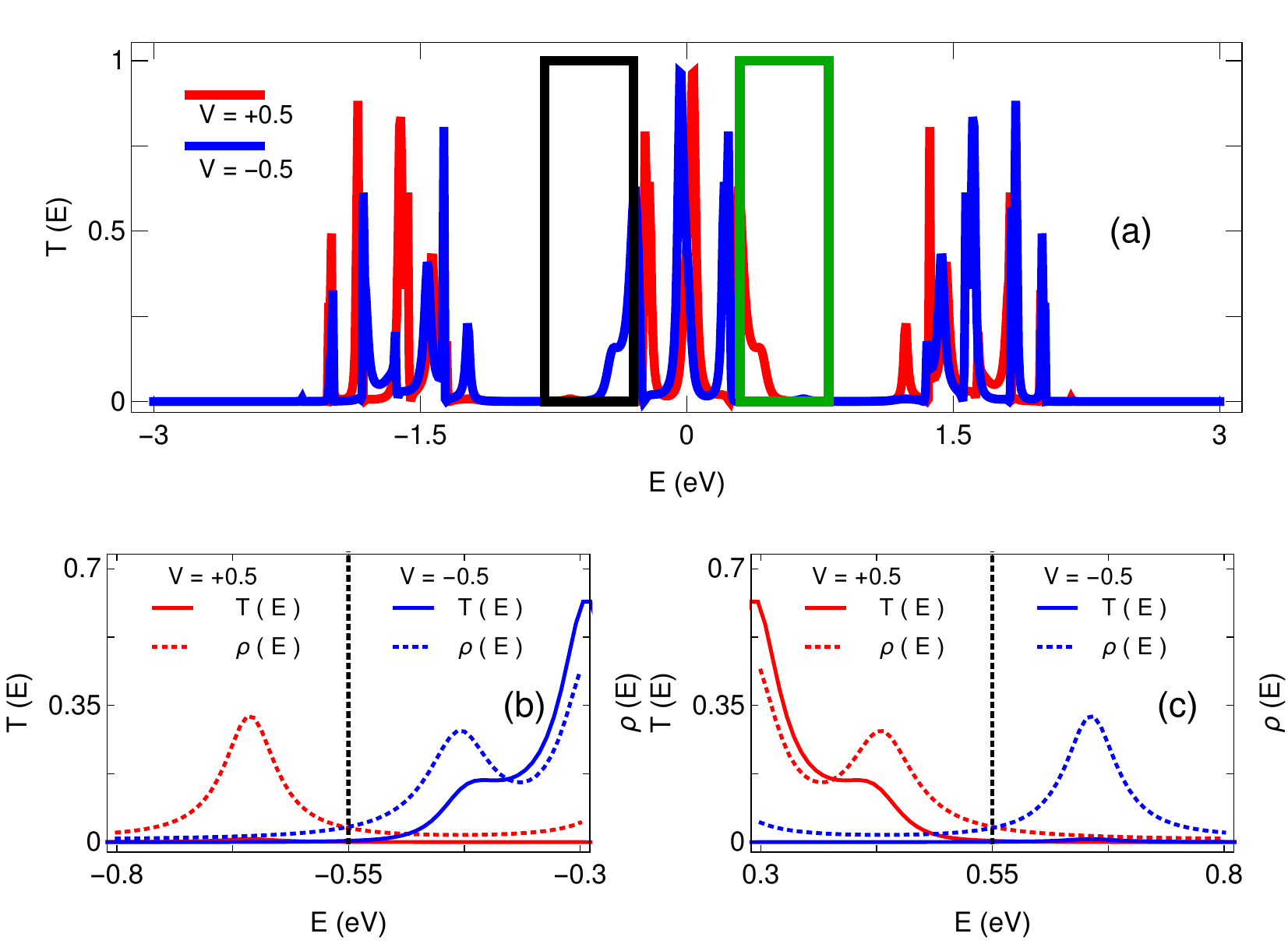}}\par}
\caption{(Color online). (a) Two-terminal transmission
probability as a function of energy for a $15$-rung ladder network under
two different bias polarities considering a typical voltage strength 
$0.5\,$V. For the two selective energy zones of (a) (marked by the black 
and green framed regions), we re-plot the transmission function in (b)
and (c) along with the ADOS for better viewing and analysis
of the results. The imaginary dotted lines in these two spectra ((b) and
(c)) represent the locations of equilibrium Fermi energy. The other 
required physical parameters for these calculations are: $t=1$,
$W_1=1$, $\phi_{\lambda}=0$, $\tau_s=1$, and $\tau_d=0.6$.}
\label{f4}
\end{figure}
Similar to the case described in the main text, let us begin with the 
results given in Fig.~\ref{f4}(a) where two-terminal transmission 
probability is
\begin{figure}[ht]
{\centering \resizebox*{8cm}{7.5cm}{\includegraphics{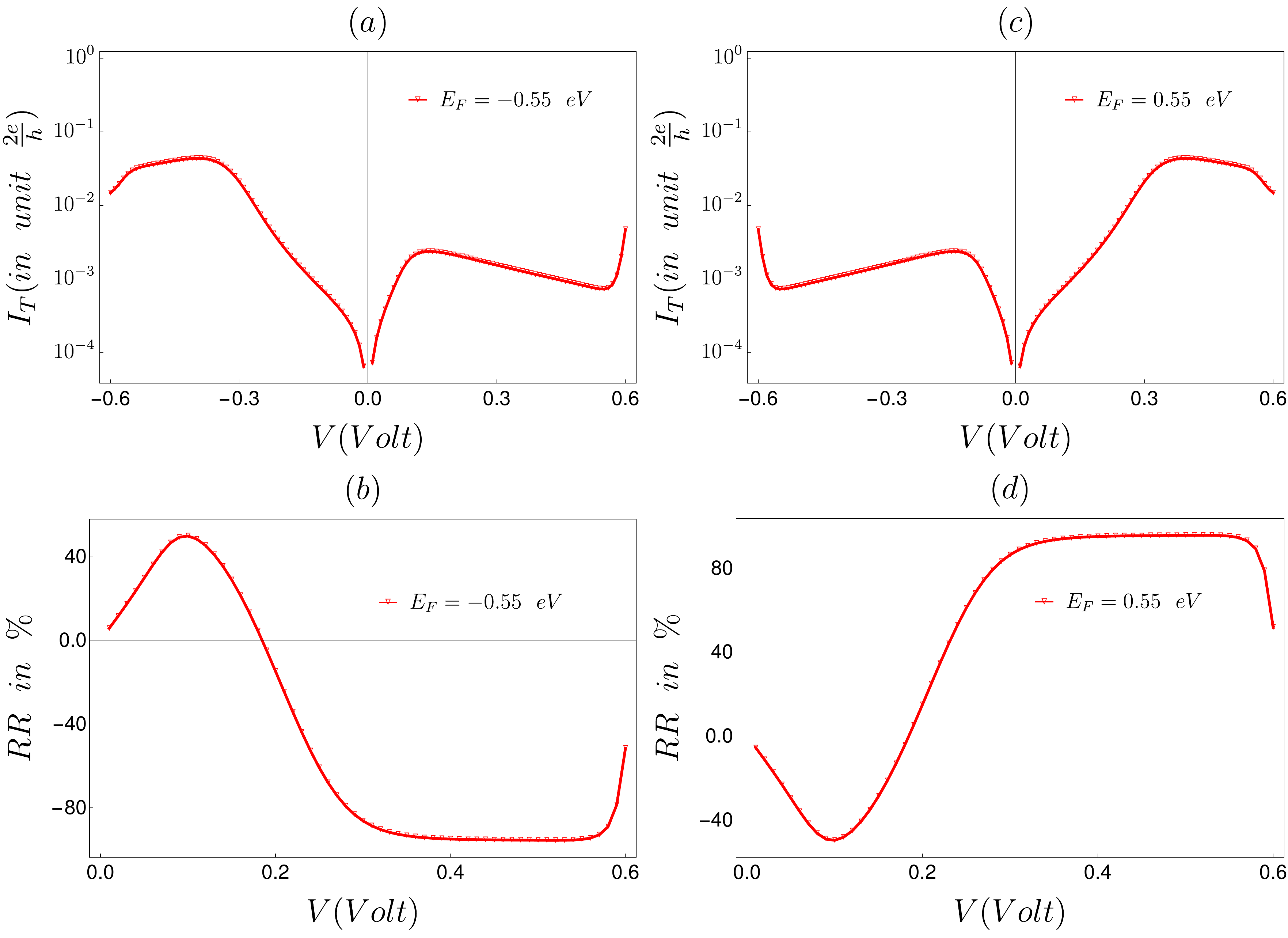}}\par}
\caption{(Color online). $I_T$-$V$ and corresponding 
$RR$-$V$ spectra at two different Fermi energies for the AAH ladder 
considering the identical physical parameters as taken in Fig.~\ref{f4}.}
\label{f5}
\end{figure}
shown for two different bias polarities considering a $15$-rung ladder
network. Sharp resonant peaks are observed at some particular energies,
and these peaks are grouped in such a way that three allowed transmitting
zones are formed and they are separated by finite gaps. This is the general 
feature of AAH type lattices, and solely associated with the ADOS spectrum. 
The other interesting feature is that the transmission function is mirror 
symmetric upon the reversal of bias polarities, that will provide another 
advantage in rectification operation as discussed below.

From the spectrum Fig.~\ref{f4}(a), we find two selective energy zones 
(marked by the black and green framed regions) of equal widths across $E=0$,
where only one colored curve dominates suppressing the other one. These 
two framed regions are redrawn in the enlarged form in Figs.~\ref{f4}(b)
and (c), where ADOS are superimposed along with the transmission
functions under two bias polarities in each of the two spectra. Clearly we
can see that for a wide energy zone ($-0.8<E<-0.3$ or $0.3<E<0.8$), either
blue curve or the red one dominates, while the other colored curve (red or
blue) almost disappears which results almost $100\%$ rectification. 
As the $T$-$E$ spectrum is mirror symmetric across $E=0$, we get equal 
magnitude of $RR$, but opposite phases setting the Fermi energy in the 
positive or negative side (as shown by the dotted vertical lines) across 
the energy band center. Comparing the transmission spectra with density of 
states, the WS localization can be noticed clearly.

Figure~\ref{f5} shows the voltage dependence of transport current and 
rectification ratio for two different choices of Fermi energy 
($E_F=-0.55\,$eV and $E_F=0.55\,$eV). Several key features are emerged.
First, the transport current is surprisingly large in one side (positive
or negative) of the bias voltage, while it is almost zero in the other side
\begin{figure}[ht]
{\centering \resizebox*{7cm}{7.5cm}{\includegraphics{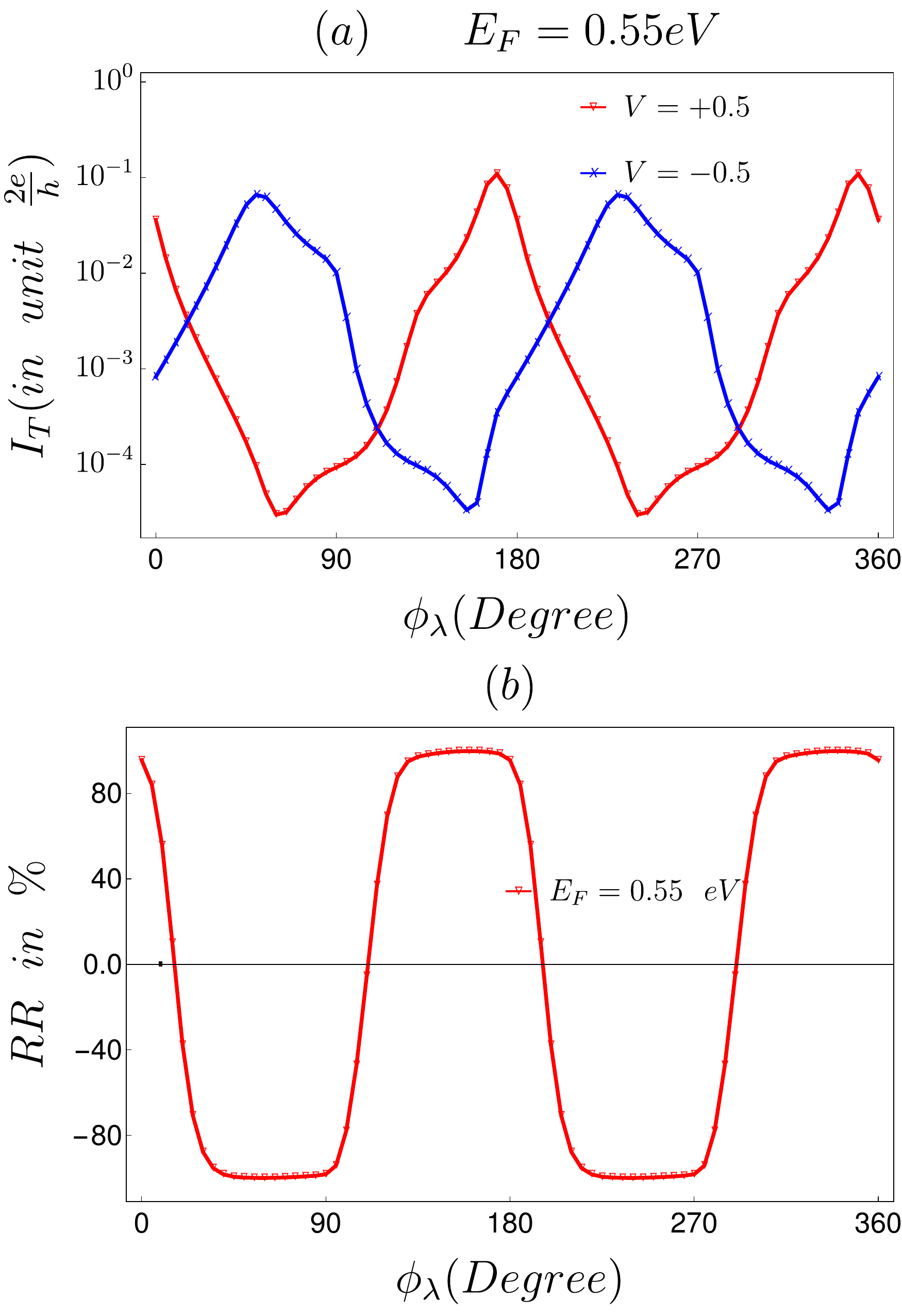}}\par}
\caption{(Color online). Role of AAH phase 
$\phi_{\lambda}$ on transport current and rectification ratio. In (a), 
transport currents are shown for two bias polarities, represented by two
different colored curves, and the corresponding $RR$ is given in
(b). Here we choose $E_F=0.55\,$eV, and all the other physical
parameters are kept unchanged as taken in Fig.~\ref{f4}.}
\label{f6}
\end{figure}
of the voltage (Figs.~\ref{f5}(a) and (c)), and this feature persists in 
a reasonable voltage range (here it is $0<V<0.6\,$V). Naturally we can 
expect a large rectification and it is clearly reflected from the spectra 
Figs.~\ref{f5}(b) and (d). Nearly $100\%$ perfect rectification is obtained
within the range $0.3<V<0.6\,$V. Second, since the $I_T$-$V$ spectrum is
mirror symmetric across $V=0$, a complete phase reversal takes place in the
$RR$-$V$ spectrum under the swapping the Fermi energy from positive zone to
the negative one. Thus, a better control of rectification is expected for
this AAH ladder network compared to the previous one in the main text.
Third, though the current initially increases with bias voltage but it 
shows a decreasing nature at higher voltages (see Figs.~\ref{f5}(a) and 
(c)). Normally for the conducting junctions where bias drop takes place
only at the junction points we get increasing current with voltage, as 
more and more contributing energy levels come within the voltage window.
But for this case since the site energies are field dependent associated 
with the applied voltage, the site energies get modified for each bias and 
accordingly ADOS, and thus, transmitting peaks are shifted. As a result, 
the number of resonant states may decrease with increasing voltage window 
\begin{figure}[ht]
{\centering \resizebox*{7.5cm}{7cm}{\includegraphics{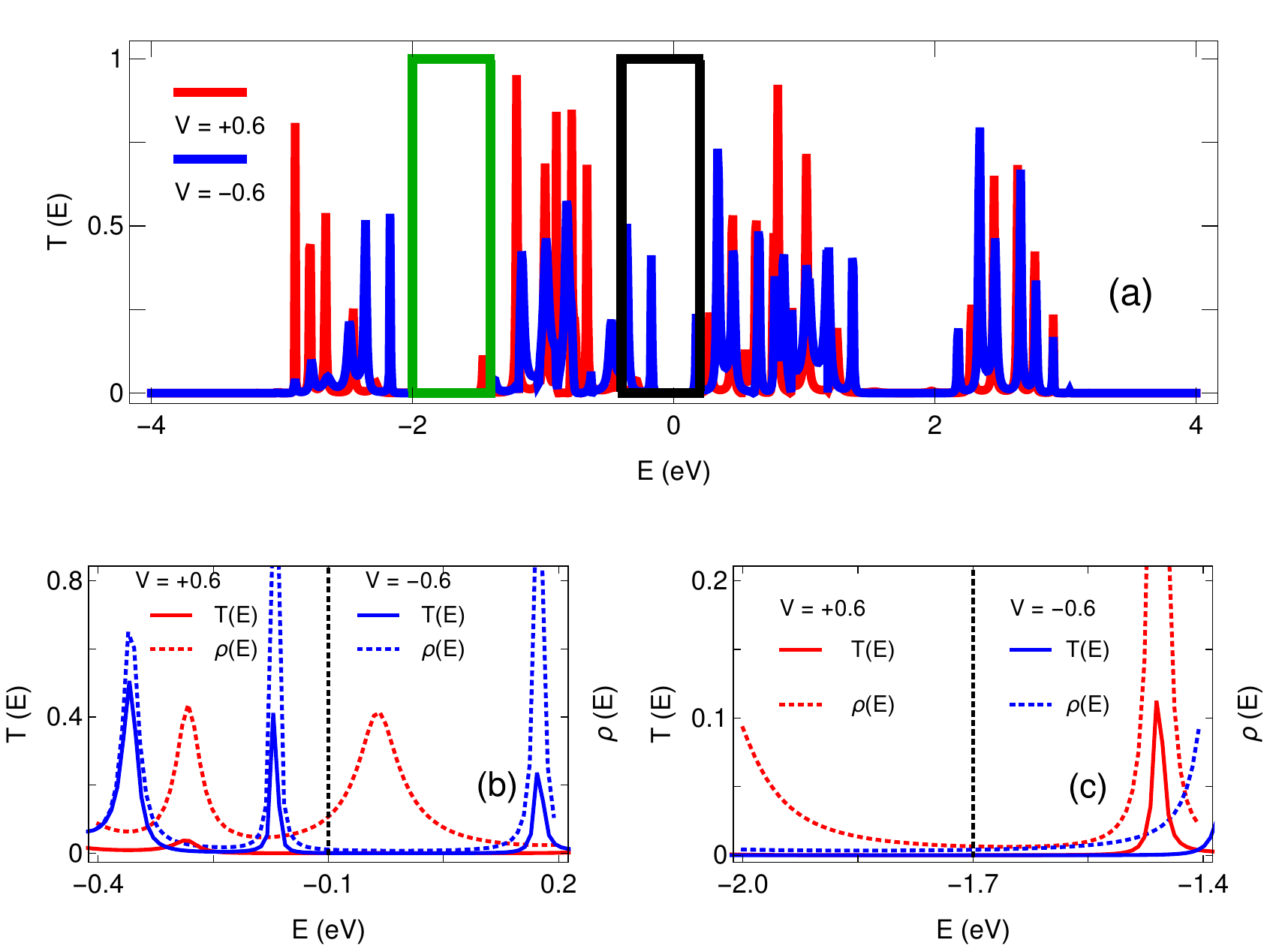}}\par}
\caption{(Color online). Same as Fig.~\ref{f4}, with
$N=21$, $W=0.8$, $W_1=0.8$, $v=1$, $t=1$, $\phi_{\lambda}=\phi_{\beta}=0$,
$\tau_s=1$, and $\tau_d=0.6$. Here we fix the voltage strength at $0.6\,$V.}
\label{f7}
\end{figure}
which yields reducing current, exhibiting the well-known negative 
differential conductance (NDC) phenomenon~\cite{ndc}.

To explore the critical role of the phase factor $\phi_{\lambda}$ on 
rectification operation for this ladder network, 
in Fig.~\ref{f6} we present the transport currents, for two bias polarities, 
along with the $RR$ as a function of $\phi_{\lambda}$. Surprisingly we see
that for a fixed (wide) $\phi_{\lambda}$ window finite current is obtained 
for one bias polarity, while the other becomes nearly zero, and this feature 
alternates with changing the phase window (see Fig.~\ref{f6}(a)). Due to this
peculiar nature of current we get almost $100\%$ rectification with 
proper phase reversal upon the variation of the AAH phase $\phi_{\lambda}$.
Thus, undoubtedly the AAH phase has a significant role on rectification 
operation.

\subsubsection{AAH modulation in the onsite energy of two strands}
\label{AAH_onsite}
\begin{figure}[ht]
{\centering \resizebox*{8cm}{7.5cm}{\includegraphics{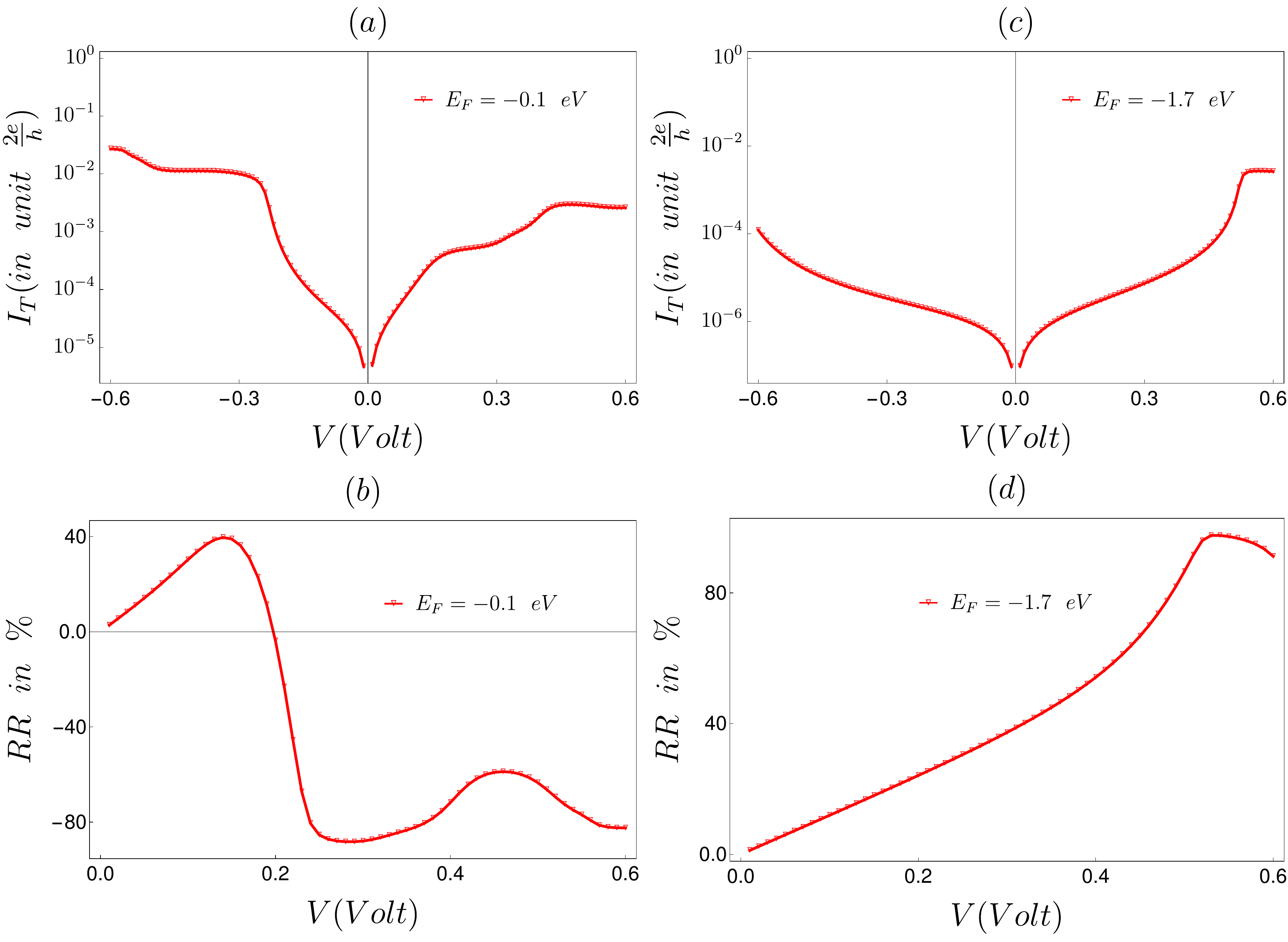}}\par}
\caption{(Color online). Different spectra represent
the identical meanings as described in Fig.~\ref{f5}. The other physical
parameters are same as used in Fig.~\ref{f7}.}
\label{f8}
\end{figure}

\begin{figure}[ht]
{\centering \resizebox*{7cm}{7.5cm}{\includegraphics{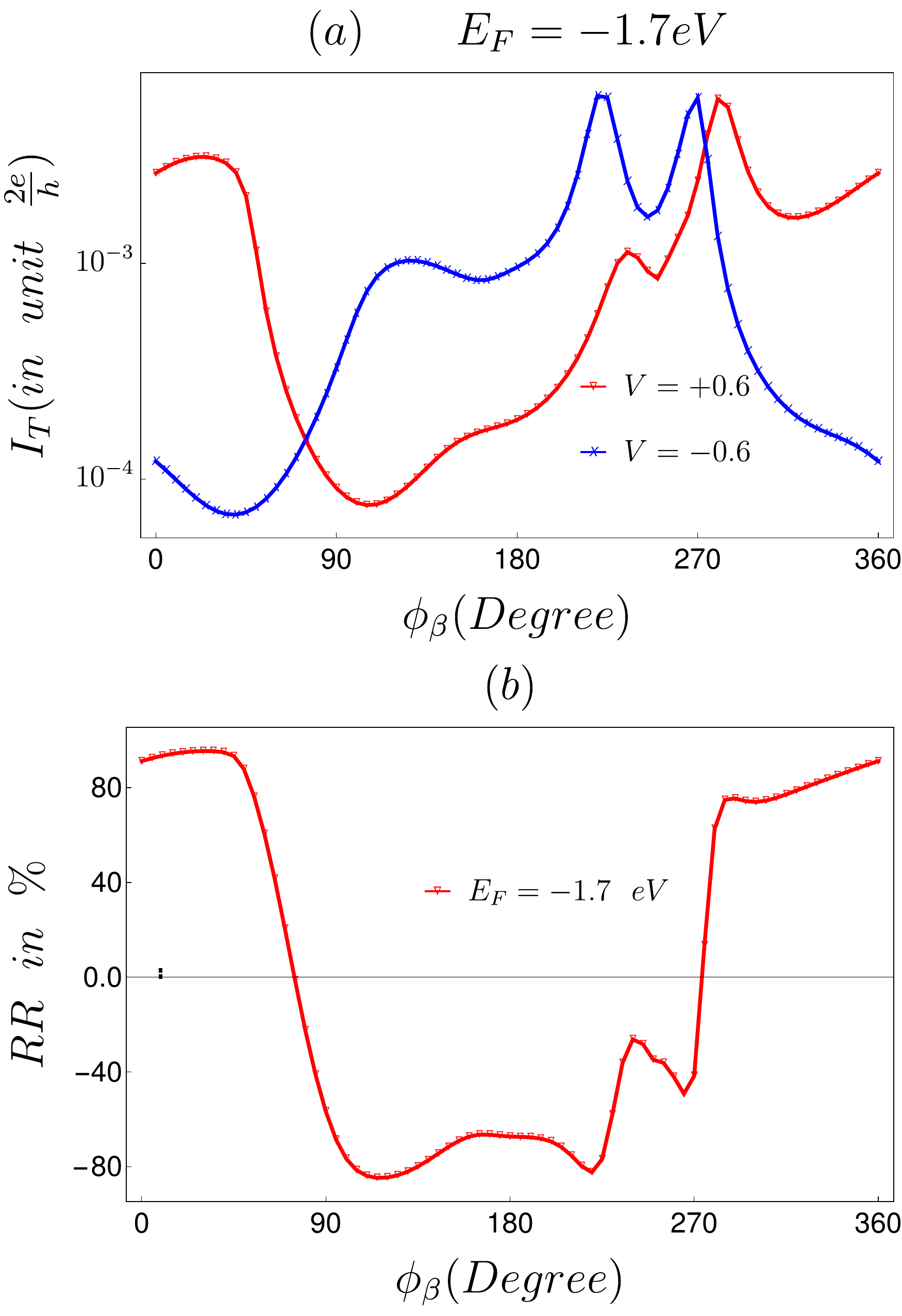}}\par}
\caption{(Color online). Upper and lower panels
represent the similar meaning as mentioned in Fig.~\ref{f3} with
$E_F=-1.7\,$eV and voltage strength $0.5\,$V. The values of the other
physical parameters are same as taken in Fig.~\ref{f7}.}
\label{f9}
\end{figure}

Here, both the two strands are modulated by AAH potentials with
different modulation strengths, $W$ and $W_1$, and they are expressed as:
$\epsilon_{Ij}^0 =W \cos\left[2\pi b j + \phi_{\beta}\right]$ and 
$\epsilon_{IIj}^0 = W_1 \cos\left[2\pi b j + \phi_{\beta}\right]$. For this
case we choose uniform intra-strand and inter-strand hopping integrals i.e.,
$t_{Ij}=t_{IIj} = t$ and $v_j = v \, \forall \,j$.
In Figs.~\ref{f7} - \ref{f9} we plot the transmission
probabilities, dependencies of junction currents and rectification ratio 
with bias voltage $V$ and phase factor $\phi_{\beta}$, like the other cases.
Without going into the detailed analysis of each of these spectra 
(Figs.~\ref{f7} - \ref{f9}) here we summarize the outcomes, as the essential
physical mechanisms behind the rectification operations are already 
discussed above. It is always possible to find suitable energy window(s) 
where transmission function for one bias polarity significantly dominates,
suppressing it in the other bias polarity (as clearly visible from the 
spectra Fig.~\ref{f7}). The complete suppression yields $100\%$
rectification, otherwise reduced rectification ratio (though its large)
is obtained (see Fig.~\ref{f8}). Two important things we need to consider 
for designing an efficient device those are: (i) the degree of $RR$ and 
(ii) the voltage region for which the high degree of rectification persists. 
For this type of ladder network (case III) we can also achieve these goals. 
The degree of rectification can be tuned as well with the help of external 
phase factor $\phi_{\beta}$ (see Fig.~\ref{f9}), though in this case 
sequential phase reversal with almost $100\%$ rectification in both 
positive and negative directions is not available like what we get in 
Fig.~\ref{f6}.

\subsubsection{AAH modulation in intra strand hopping}
\label{AAH_intra}
Finally, we consider the case where the AAH modulation is introduced in 
the intra-strand hopping integrals, keeping uniform hopping 
\begin{figure}[ht]
{\centering \resizebox*{7.5cm}{7.5cm}{\includegraphics{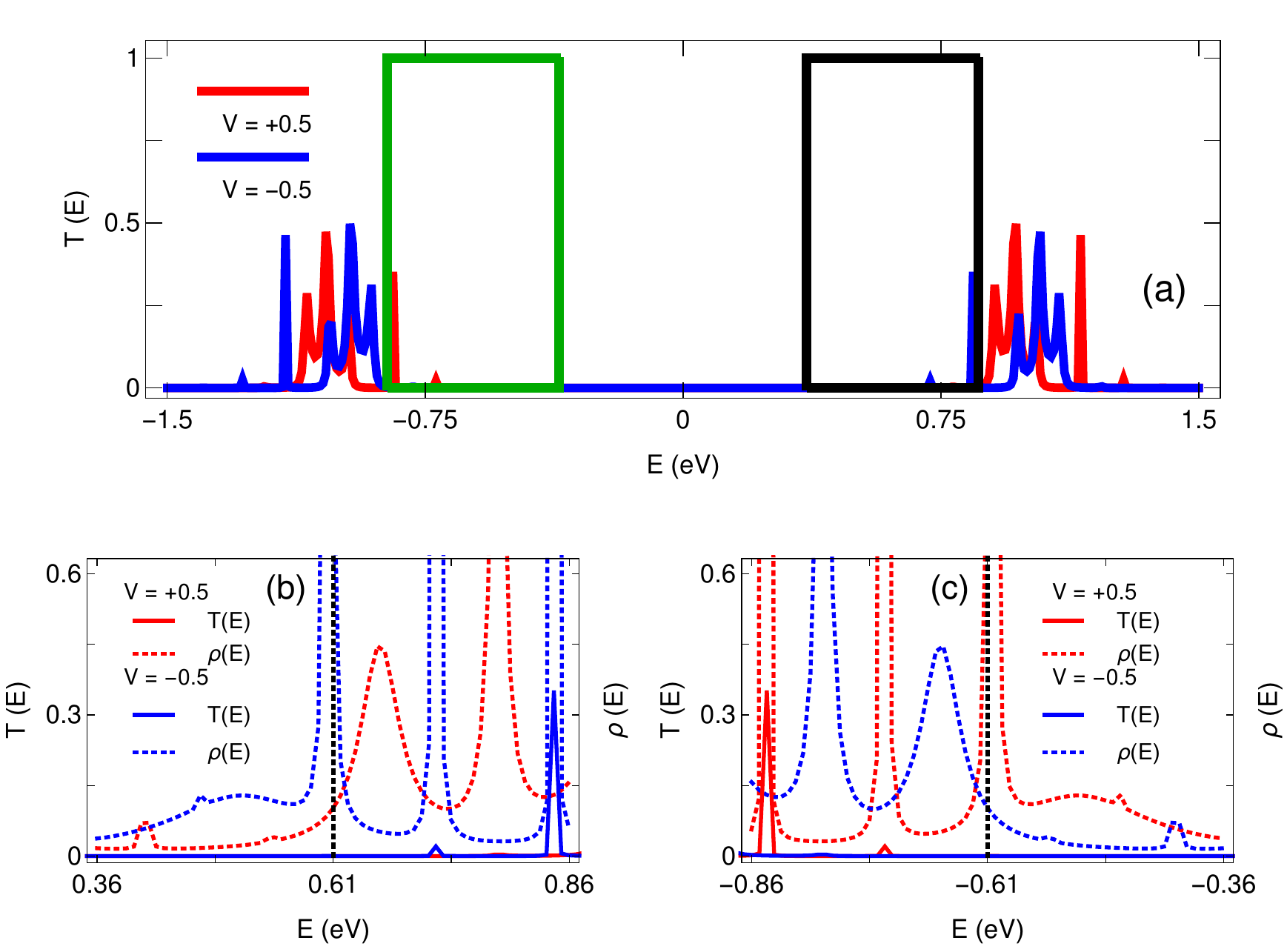}}\par}
\caption{(Color online). Same as Fig.~\ref{f4}, with
$N=15$, $\epsilon_{Ij}^0=\epsilon_{IIj}^0=0$, $W=0.4$, $W_1=0.4$, $v=1$,
$\phi_{\lambda}=\phi_{\beta}=0$, $\tau_s=1$, and $\tau_d=0.6$.}
\label{f10}
\end{figure}
\begin{figure}[ht]
{\centering \resizebox*{8cm}{7.5cm}{\includegraphics{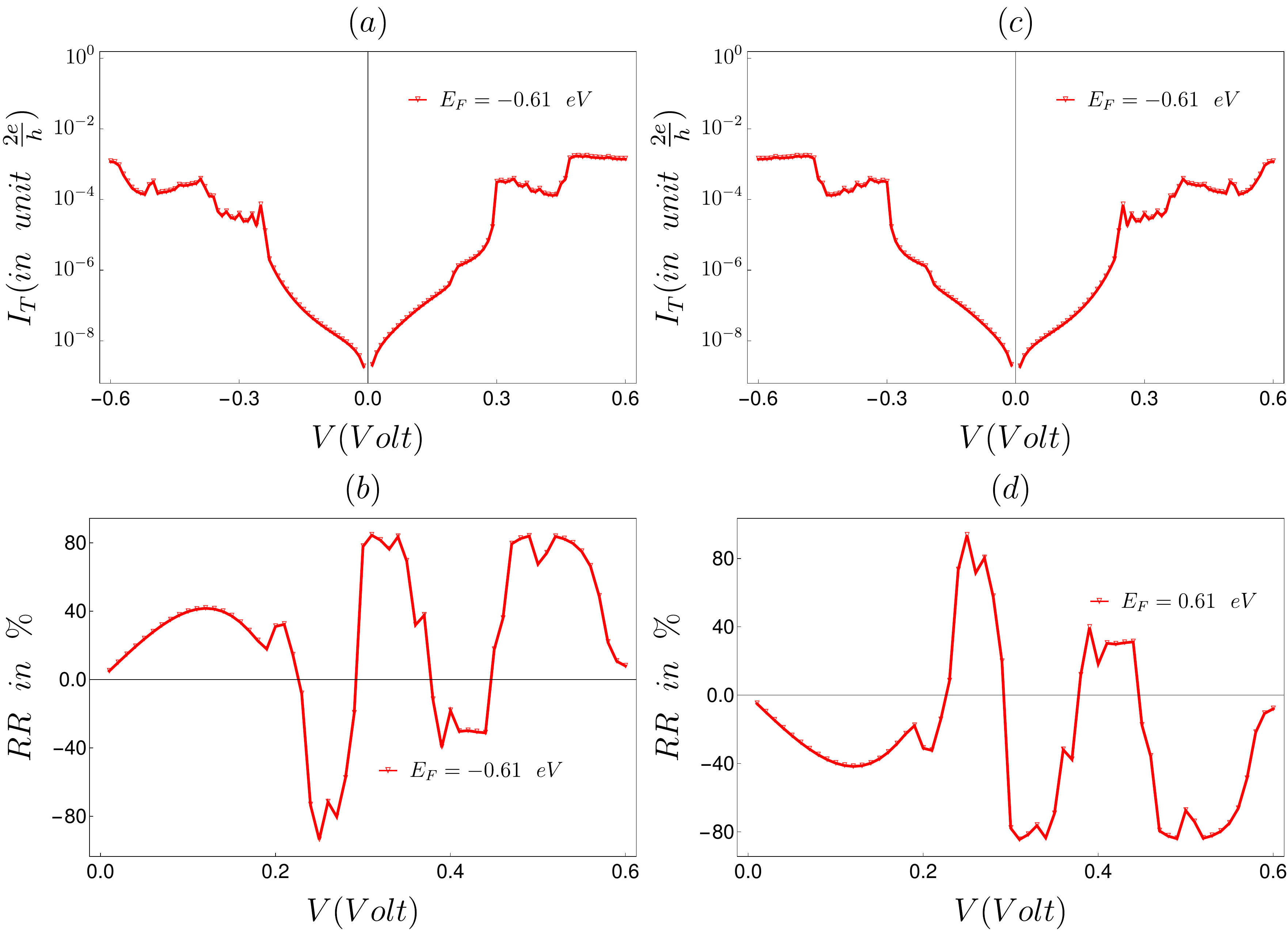}}\par}
\caption{(Color online). Different spectra represent
the identical meanings as described in Fig.~\ref{f8}. The other physical
parameters are same as used in Fig.~\ref{f10}.}
\label{f11}
\end{figure}
between the two strands (viz, $v_j =v \, \forall \,j$). Thus, for this case 
the intra-strand hopping integrals are described as
$t_{Ij} = W \cos\left[2\pi b j + \phi_{\lambda}\right]$ and
$t_{IIj} = W_1 \cos\left[2\pi b j + \phi_{\lambda}\right]$. Here, bias
independent site energies are also constant and put them to zero i.e.,
$\epsilon_{Ij}^0=0$ and $\epsilon_{IIj}^0=0$.

\begin{figure}[ht]
{\centering \resizebox*{7cm}{7.5cm}{\includegraphics{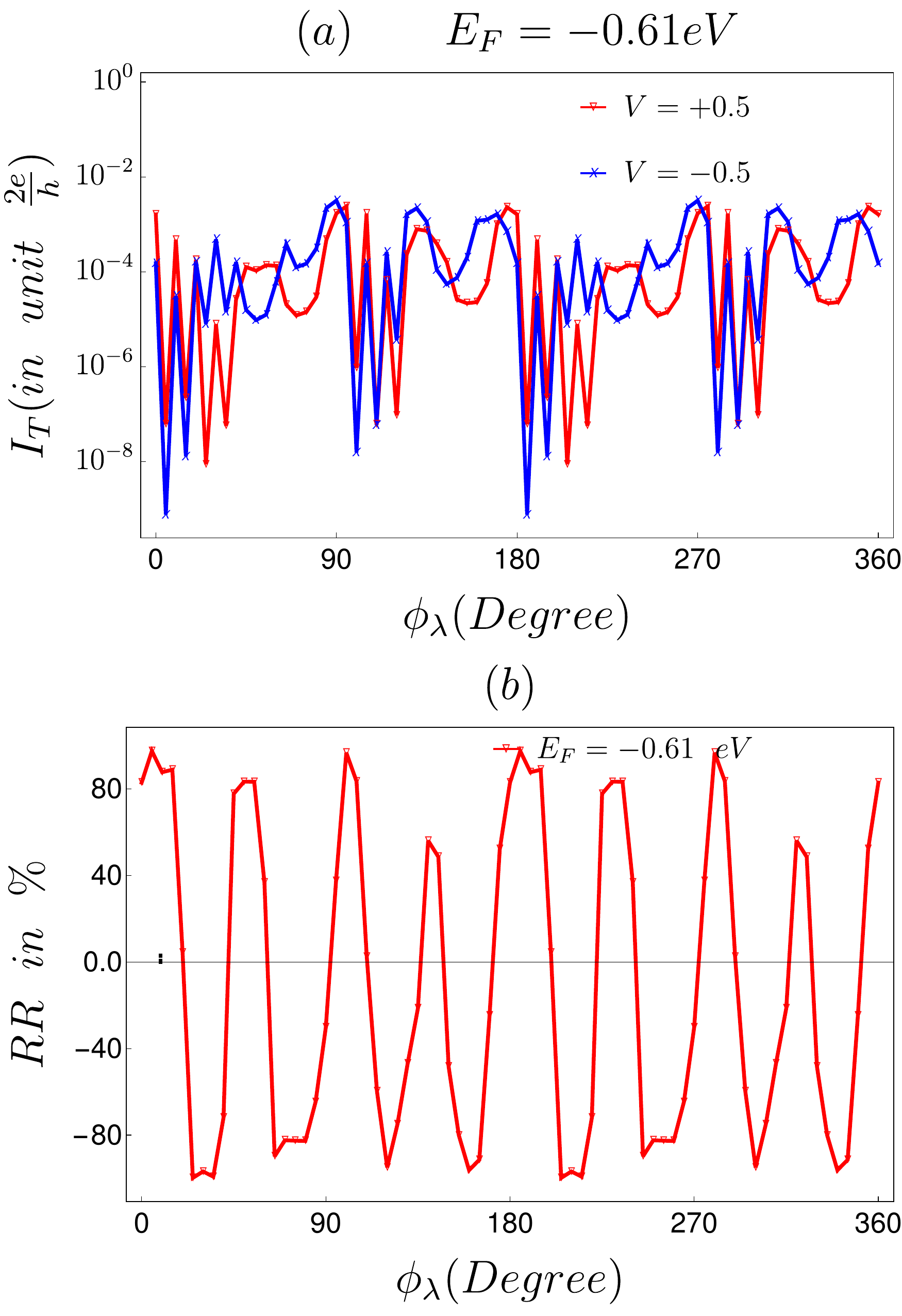}}\par}
\caption{(Color online). Upper and lower panels
represent the similar meaning as described in Fig.~\ref{f6} with
$E_F=-0.61\,$eV. The values of the other physical parameters are same
as taken in Fig.~\ref{f10}.}
\label{f12}
\end{figure}

Let us focus on the spectra given in Figs.~\ref{f10} - \ref{f12} where 
the results are shown for the AAH ladder described in this subsection.
In this type of ladder, the intra-strand hopping integrals in  
both the upper and lower strands are modulated with AAH distributions having 
the strengths $W$ and $W_1$, respectively, associated with the phase factor
$\phi_{\lambda}$. The resonant peaks are packed together forming quasi-band
like structures, and they are separated far away from each other (see 
Fig.~\ref{f10}(a)). Here also we can find some energy zones, like other 
cases, where one of the resonant curves associated with a particular bias
polarity dominates over the other (Figs.~\ref{f10}(b) and (c)), which 
demonstrates the possibilities of getting rectification action. For this
setup the transport currents in positive and negative biases are quite
low and nearly equal in the low bias region (see Figs.~\ref{f11}(a) and (c))
resulting smaller degree of current rectification. Whereas, reasonably 
large degree of rectification is achieved in the limit of higher voltage
($V>\sim 0.25\,$V) associated with the $I_T$-$V$ characteristics.
Setting into this moderate voltage region if we tune the phase factor 
$\phi_{\lambda}$, choosing a suitable Fermi energy, we can see that
several phase windows (relatively small widths) are available where 
transport current for one bias polarity dominates over the other, and 
the dominating nature i.e., which curve dominates the other depends on
the phase factor (see Fig.~\ref{f12}(a)). As a result of this, nice 
oscillating pattern is generated in RR upon the variation of 
$\phi_{\lambda}$ (Fig.~\ref{f12}(b)), providing high degree of 
rectification. 

\section{Combined effect of $\phi_{\beta}$ and $\phi_{\lambda}$ on
rectification}
\label{bb}

The results described above in different configurations
initiate the obvious curiosity that how the results of rectification depend
\begin{figure}[ht]
\centering
\includegraphics[width=0.5\textwidth]{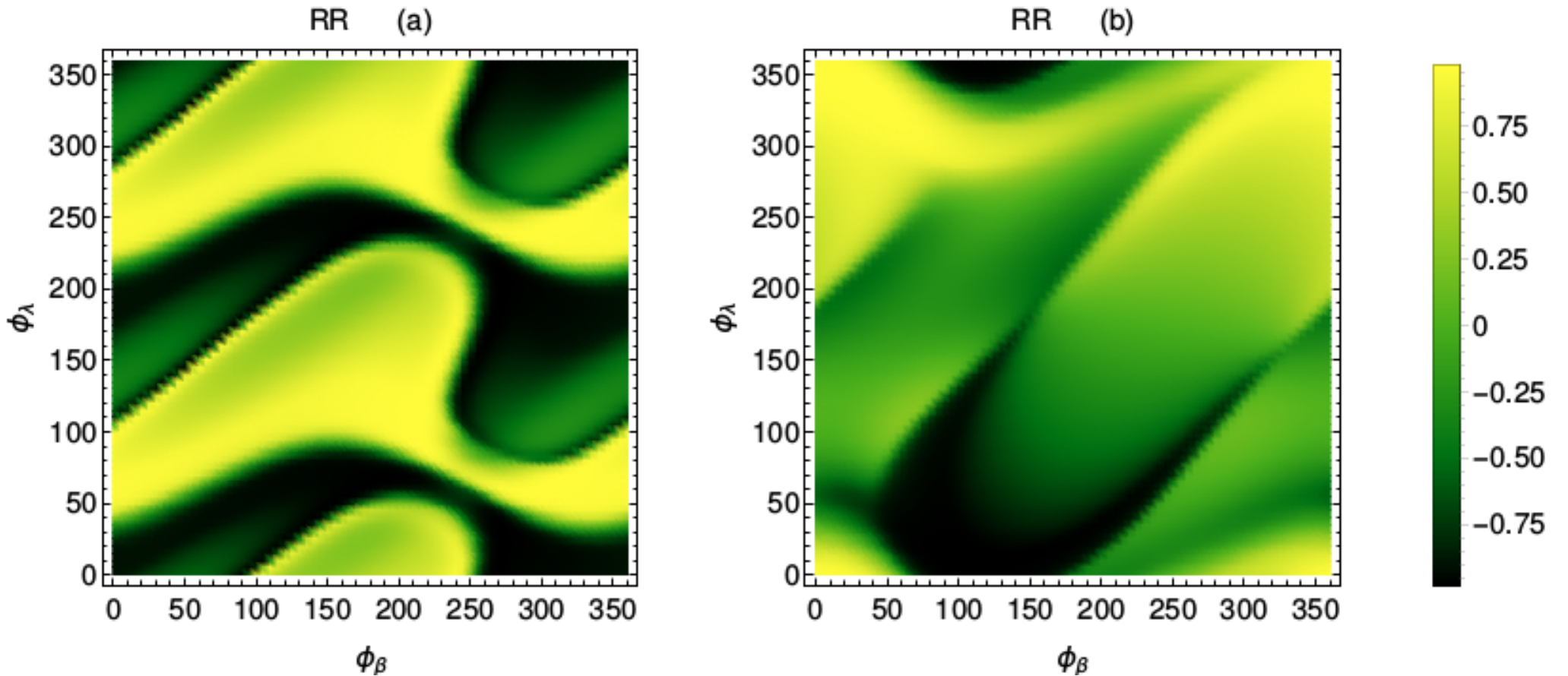}
\caption{(Color online). $RR$ as functions of $\phi_{\lambda}$ and
$\phi_{\beta}$ for two typical ladders. (a) $\epsilon_{Ij}^0=W
\cos[2\pi b j + \phi_{\beta}]$ and $\epsilon_{IIj}^0=W_1
\cos[2\pi b j + \phi_{\beta}]$; $v_j=v_0 \cos[2\pi b j + \phi_{\lambda}]$;
$t_{Ij}=t_{Ij}=t\, \forall \,j$, and
(b) $\epsilon_{Ij}^0=W \cos[2\pi b j + \phi_{\beta}]$ and
$\epsilon_{IIj}^0=W_1 \cos[2\pi b j + \phi_{\beta}]$; $t_{Ij}=t_{IIj} =
t_0 \cos[2\pi b j + \phi_{\lambda}]$; $v_j=v\, \forall \,j$. For the
calculations we choose (a) $W=W_1=0.5$, $v_0=1$, $t=1$, $E_F=-0.7$ and
(b) $W=W_1=0.4$, $v_0=1$, $t_0=0.4$, $E_F=1.5$. The other parameter values
are: $N=13$ and bias voltage $V=0.4\,$V.}
\label{f14}
\end{figure}
on the two phases if we vary them together. To unravel it, in Fig.~\ref{f14}
we plot (density plot) the $RR$s for two typical AAH ladders as functions of
$\phi_{\lambda}$ and $\phi_{\beta}$. These ladders are quite different than
the previously mentioned cases, as here we consider
$\phi_{\lambda}$ along with $\phi_{\beta}$ in the Hamiltonian. The two
ladder systems are specified as follows: in one case (the results of which
are shown in Fig.~\ref{f14}(a)) the upper and lower strands are subjected
to AAH potentials having the
strengths $W$ and $W_1$, respectively, with the phase modulation factor 
$\phi_{\beta}$. Along with this, the AAH modulation is also introduced in 
the inter-strand hopping integrals with unit strength, associated with the 
phase factor $\phi_{\lambda}$, keeping the intra-strand hopping integrals
uniform. For the other ladder system (results
of which are given in Fig.~\ref{f14}(b)), the two strands together with 
intra-strand hopping integrals are modulated in the AAH form, keeping the
inter-strand hopping integrals as uniform. The results in both these two
cases are quite interesting as we can see that for a wide range of the 
phase factors almost $100\%$ rectification is available and with 
changing these phases the sign alteration also takes place sequentially 
providing the same degree of rectification.


\begin{thebibliography}{99}

\bibitem{pr1} A. Nitzan and M. A. Ratner, Science \textbf{300}, 1384 (2003).

\bibitem{pr2} N. J. Tao, Nature Nanotech. \textbf{1}, 173 (2006).

\bibitem{pr3} S. V. Aradhya and L. Venkataraman, Nature Nanotech. 
\textbf{8}, 399 (2013).

\bibitem{pr4} A. Aviram and M. A. Ratner, Chem. Phys. Lett. \textbf{29},
277 (1974).

\bibitem{pr5} P. E. Kornilovitch, A. M. Bratkovsky, and R. S. Williams,
Phys. Rev. B \textbf{66}, 165436 (2002).

\bibitem{pr6} J. Taylor, M. Brandbyge, and K. Stokbro, Phys. Rev. Lett. 
\textbf{89}, 138301 (2002).

\bibitem{pr7} M. Paulsson, F. Zahid, and S. Datta, {\em Handbook of 
Nanoscience, Engineering, and Technology} (2003).

\bibitem{pr8} R. M. Metzger, Acc. Chem. Res. \textbf{32}, 950 (1999).

\bibitem{pr9} M. Elbing, R. Ochs, M. Koentopp, M. Fischer, C. von Hanisch,
F. Weigend, F. Evers, H. B. Weber, and M. Mayor, Proc. Natl. Acad. Sci. 
U.S.A. \textbf{102}, 8815 (2005).

\bibitem{pr10} I. Diez-Perez, J. Hihath, Y. Lee, L. Yu, L. Adamska, M. A.
Kozhushner, I. I. Oleynik, and N. Tao, Nature Chemistry \textbf{1}, 635 
(2009).

\bibitem{pr11} J. Hihath, C. Bruot, H. Nakamura, Y. Asai, I. Diez-Perez,
Y. Lee, L. Yu, and N. Tao, ACS Nano \textbf{5}, 8331 (2011).

\bibitem{pr12} E. L{\"o}rtscher, B. Gotsmann, Y. Lee, L. Yu, C. Rettner, 
and H. Riel, ACS Nano \textbf{6}, 4931 (2012).

\bibitem{pr13} A. Batra, P. Darancet, Q. Chen, J. S. Meisner, J. R. Widawsky,
J. B. Neaton, C. Nuckolls, and L. Venkataraman, Nano Lett. \textbf{13}, 
6233 (2013).

\bibitem{pr14} T. Kim, Z. F. Liu, C. Lee, J. Neaton, and L. Venkataraman, 
Proc. Natl. Acad. Sci. U. S. A. \textbf{111}, 10928 (2014).

\bibitem{pr15} A. Batra, J. S. Meisner, P. Darancet, Q. Chen, C. 
Steigerwald, M. L. Nuckolls, and L. Venkataraman, Faraday Discussions 
\textbf{174}, 79 (2014).

\bibitem{pr16} B. Capozzi, J. Xia, O. Adak, E. J. Dell, Z.-F. Liu, 
J. C. Taylor, J. B. Neaton, L. M. Campos, and L. Venkataraman, 
Nat. Nanotechnol. \textbf{10}, 522 (2015).

\bibitem{pr17} A. Dhirani, P. -H. Lin, P. Guyot-Sionnest, R. W. Zehner, 
and L. R. Sita, J. Chem. Phys. \textbf{106}, 5249 (1997).

\bibitem{pr18} C. Zhou, M. R. Deshpande, M. A. Reed, L. Jones, and 
J. M. Tour, Appl. Phys. Lett. \textbf{71}, 611 (1997).

\bibitem{pr19} J. Frantti, V. Lantto, S. Nishio, and M. Kakihana, Phys. 
Rev. B \textbf{59}, 12 (1999).

\bibitem{pr20} G. J. Ashwell, W. D. Tyrrell, and A. J. Whittam, J. Am. Chem. 
Soc. \textbf{126}, 7102 (2004).

\bibitem{pr21} C. A. Nijhuis, W. F. Reus, and G. M. Whitesides, J. Am. 
Chem. Soc. \textbf{132}, 18386 (2010).

\bibitem{pr22} S. K. Yee {\em et al.}, ACS Nano \textbf{5}, 9256 (2011).

\bibitem{pr23} H. J. Yoon {\em et al.}, J. Am. Chem. Soc. \textbf{136}, 
17155 (2014).

\bibitem{pr24} K. Wang, J. Zhou, J. M. Hamill, and B. Xu, J. Chem. Phys. 
\textbf{141}, 054712 (2014).

\bibitem{pr25} B. Capozzi {\em et al.}, Nano Lett. \textbf{14}, 1400 (2014).

\bibitem{pr26} M. L. Perrin, E. Gal{\'a}n, R. Eelkema, J. M. Thijssen, 
F. Grozema, and  H. S. J. van der Zant, Nanoscale \textbf{8}, 8919 (2016).

\bibitem{pr27} X. Chen, M. Roemer, L. Yuan, W. Du, D. Thompson, E. D. Barco, 
and C. A. Nijhuis, Nat. Nanotech. \textbf{12}, 797 (2017).

\bibitem{pr28} J. T. Obodo, A. Murat and U. Schwingenschl{\"o}gl, 
Sci. Report \textbf{7}, 7324 (2017).

\bibitem{dna1} M. D. Ventra and Y. V. Pershin, Nature Nanotech. \textbf{6}, 
198 (2011).   
 
\bibitem{dna2} J. C. Genereux and J. K. Barton, Nature Chem. \textbf{1}, 
106 (2009). 

\bibitem{dna3} G. I. Livshits {\em et al.}, Nature Nanotech. \textbf{9}, 
1040 (2014).

\bibitem{dna4} C. Dekker and M. A. Ratner, Phys. World \textbf{14}, 29 (2001).

\bibitem{dna5} A. Guo and Q.-F. Sun, Phys. Rev. B \textbf{86}, 115441 (2012).

\bibitem{dna6} C. Guo {\em et al.}, Nature Chem. \textbf{8}, 484 (2016).

\bibitem{quasiall} A. M. Guo, Phys. Rev. E \textbf{75}, 061915 (2007).

\bibitem{tm} H. Lei, J. Chen, G. Nouet, S. Feng, Q. Gong, and X. Jiang,
Phys. Rev. B \textbf{75}, 205109 (2007).

\bibitem{quasi-recti} M. Saha and S. K. Maiti, Physica E \textbf{93}, 275 
(2017).

\bibitem{macia06} E. Maci\'{a}, Phys. Rev. B \textbf{74}, 245105 (2006).

\bibitem{rudo07} G. Cuniberti, E. Maci\'{a}, A. Rodr\'{i}guez, and
R. A. R\"{o}mer, in {\em Charge Migration in DNA: Perspectives from 
Physics, Chemistry and Biology} edited by T. Chakraborty, Springer-Verlag, 
Berlin (2007).

\bibitem{ladder1} S. Sil, S. K. Maiti, and A. Chakraborti, Phys. Rev. Lett.
\textbf{101}, 076803 (2008).

\bibitem{ladder2} S. Sil, S. K. Maiti, and A. Chakrabarti, Phys. Rev. B 
\textbf{78}, 113103 (2008). 

\bibitem{moura} E. L. Albuquerqu, U. L. Fulco, V. N. Freire, E. W. S. 
Caetano, M. L. Lyra, and F. A. B. F. deMoura, Phys. Rep. \textbf{535},
139 (2014).

\bibitem{sousa} L. M. Bezerril, D. A. Moreir, E. L. Albuquerque, U. L. 
Fulco, E. L. de Oliveira, and J. S. de Sousa, Phys. Lett. A \textbf{373}, 
3381 (2009).

\bibitem{ladder3} S. Maiti, S. Sil, and A. Chakrabarti, Ann. Phys. (N. Y.)
\textbf{382}, 150 (2017).

\bibitem{aubry1} S. Aubry and G. Andre, Ann. Isr. Phys. Soc. \textbf{3}, 
133 (1980).

\bibitem{aubry2} Y. E. Kraus, Y. Lahini, Z. Ringel, M. Verbin, and 
O. Zilberberg, Phys. Rev. Lett. \textbf{109}, 106402 (2012).

\bibitem{aubry3} M. Verbin, O. Zilberberg, Y. E. Kraus, Y. Lahini, and Y.
Silberberg, Phys. Rev. Lett \textbf{110}, 076403 (2013).

\bibitem{expt0} M. Schreiber, S. S. Hodgman, P. Bordia, H. P. L{\"u}schen, M. H. Fischer , R.
Vosk, E. Altman, U. Schneider, and I. Bloch, Science \textbf{349}, 842 (2015).

\bibitem{expt1} H. P. L{\"u}schen, S. Scherg, T. Kohlert, M. Schreiber, 
P. Bordia, X. Li, S. D. Sarma, and I. Bloch, Phys. Rev. Lett. \textbf{120}, 
160404 (2018).

\bibitem{expt2} H. P. L{\"u}schen, P. Bordia, S. S. Hodgman, M. Schreiber, 
S. Sarkar, A. J. Daley, M. H. Fischer, E. Altman, Immanuel Bloch, and 
U. Schneider, Phys. Rev. X. \textbf{7}, 011034 (2017).

\bibitem{expt3} M. Atala, M. Aidelsburger, M. Lohse, J. T. Barreiro, 
B. Paredes, I. Bloch, Nature Phys. \textbf{10}, 588 (2014).

\bibitem{aubry4} A. Purkayastha, S. Sanyal, A. Dhar, and M. Kulkarni, 
Phys. Rev. B \textbf{97}, 174206 (2018).

\bibitem{aubry5} S. Ganeshan, K. San, and S. Das Sarma, Phys. Rev. Lett.
\textbf{110}, 180403 (2013).

\bibitem{aubry6} S. Ganeshan, J. H. Pixley, and S. Das Sarma, Phys. Rev.
Lett. \textbf{114}, 146601 (2015).

\bibitem{aubry7} A. Purkayastha, A. Dhar, and M. Kulkarni, Phys. Rev. B 
\textbf{96}, 180204 (2017).

\bibitem{aubry2d} Y. E. Kraus, Z. Ringel, and O. Zilberberg, Phys. Rev. 
Lett. \textbf{111}, 226401 (2013).

\bibitem{aubry2d2} O. Zilberberg, S. Huang, J. Guglielmon, M. Wang,
K. Chen, Y. E. Kraus, and M. C. Rechtsman, Nature \textbf{553}, 59 (2018).

\bibitem{aubry2d3} M. Lohse, C. Schweizer, H. M. Price, O. Zilberberg, and 
I. Bloch, Nature \textbf{553}, 55 (2018).

\bibitem{fragment1} M. Kohmoto, B. Sutherland, and C. Tang, Phys. Rev. B 
\textbf{35}, 1020 (1987).

\bibitem{fragment2} G. J. Jin, Z. D. Wang, A. Hu, and S. S. Jiang, 
Phys. Rev. B \textbf{55}, 9302 (1997).
 
\bibitem{quasirec1} V. Balachandran, S. R. Clark, J. Goold, and D. Poletti, 
Phys. Rev. Lett \textbf{123}, 020603 (2019).

\bibitem{green1} D. S. Fisher and P. A. Lee, Phys. Rev. B {\bf 23}, 6851 
(1981).

\bibitem{green2} S. Datta, {\it Electronic Transport in Mesoscopic Systems},
Cambridge University Press, Cambridge (1997).

\bibitem{green3} S. Datta, {\it Quantum Transport: Atom to Transistor}, 
Cambridge University Press, Cambridge (2005).

\bibitem{green4} B. K. Nikoli{\' c} and P. B. Allen, J. Phys.: Condens. Matter
\textbf{12}, 9629 (2000).

\bibitem{green5} P. Dutta, S. K. Maiti, and S. N. Karmakar, J. Appl. Phys.
\textbf{114}, 034306 (2013).

\bibitem{green6} M. Dey, S. K. Maiti, and S. N. Karmakar, Org. Electron.
\textbf{12}, 1017 (2011). 

\bibitem{inter1} S. K. Pati, J. Chem. Phys. {\bf 118}, 6529 (2003).

\bibitem{inter2} V. Mujica, M. Kemp, A. Roitberg, and M. Ratner, 
J. Chem. Phys. {\bf 104}, 7296 (1996).

\bibitem{inter3} K. Walczak, Phys. B {\bf 365}, 193 (2005).

\bibitem{inter4} A. Dey, D. S. Bhakuni, B. K. Agarwalla, and A. Sharma, 
arXiv:1902.00474v1.

\bibitem{el1} S. Pleutin, H. Grabert, G. L. Ingold, and A. Nitzan,
J. Chem. Phys. \textbf{118}, 3756 (2003).

\bibitem{el2} S. K. Maiti and A. Nitzan, Phys. Lett. A \textbf{377}, 1205 
(2013). 

\bibitem{el3} P. Dutta, S. K. Maiti, and S. N. Karmakar, AIP Advances
\textbf{4}, 097126 (2014).

\bibitem{wannier1} G. H. Wannier, Phys. Rev. {\bf 117}, 432 (1960).

\bibitem{wannier2} J. R. Borysowicz, Phys. Lett. A {\bf 231}, 240 (1997).

\bibitem{wannier3} N. Zekri, M. Schreiber, R. Ouasti, R. Bouamrane, and A.
Brezini, Z. Phys. B \textbf{99}, 381 (1996).

\bibitem{cb1} I. L. Aleiner, P. W. Brouwer, and L. I. Glazman, Phys. Rep. 
\textbf{358}, 309 (2002).

\bibitem{cb2} C. W. J. Beenakker, Phys. Rev. B \textbf{44}, 1646 (1991).

\bibitem{cb3} M. Galperin, A. Nitzan, and M. A. Ratner, Phys. Rev. B 
\textbf{78}, 125320 (2008).

\bibitem{cd1} T. J. Gramila, J. P. Eisenstein, A. H. MacDonald, L. N. 
Pfeiffer, and K. W. West, Phys. Rev. Lett. \textbf{66}, 1216 (1991).

\bibitem{cd2} U. Sivan, P. M. Solomon, and H. Shtrikman, Phys. Rev. Lett. 
\textbf{68}, 1196 (1992).

\bibitem{cd3} L. Zheng and A. H. MacDonald, Phys. Rev. B \textbf{48}, 8203 
(1993).

\bibitem{ndc} B. Xu and Y. Dubi, J. Phys.: Condens. Matter \textbf{27},
263202 (2015).

\end{thebibliography}
\end{document}